\documentclass[12pt,aps,prd,preprint,tightenlines,superscriptaddress,
   showpacs,nofootinbib]{revtex4}
\newcommand{\PRE}[1]{{#1}} % Use if preprint style
\usepackage{bm} 
\usepackage{epsfig}

\newcommand{\postscript}[2]{\setlength{\epsfxsize}{#2\hsize}
\centerline{\epsfbox{#1}}}

\newcommand{\OmegaSWIMP}{\Omega_{\text{SWIMP}}}

\newcommand{\kev}{\text{keV}} \newcommand{\mev}{\text{MeV}}
\newcommand{\gev}{\text{GeV}} \newcommand{\tev}{\text{TeV}}
 \newcommand{\mb}{\text{mb}}
 
 \newcommand{\s}{\text{s}}

\newcommand{\eqref}[1]{Eq.~(\ref{#1})}

\newcommand{\Gravitino}{\tilde{G}}

\newcommand{\mgravitino}{m_{\gravitino}}
\newcommand{\gravitino}{\tilde{G}}

\def\Real{\Re e\,}
\def\Imag{\Im m\,}
\def\mg{m_{\tilde{G}}}
\def\msl{m_{\tilde{l}}}
\def\pstau{p_{\tilde{l}}}

\def\pZ{p_V}
\def\pg{p_{\tilde{G}}}
\def\pgpZ{\pg\cdot\pZ}
\def\pgpsl{\pg\cdot\pstau}
\def\pslpZ{\pstau\cdot\pZ}

\def\calO{{\cal{O}}}

\begin{document}

\preprint{UCI-TR-2004-4}  \preprint{hep-ph/0404198}

\title{ \PRE{\vspace*{1.5in}}
SuperWIMP Gravitino Dark Matter from\\
Slepton and Sneutrino Decays 
\PRE{\vspace*{0.3in}} }

\author{Jonathan L.~Feng}
\affiliation{Department of Physics and Astronomy, University of
California, Irvine, CA 92697, USA
\PRE{\vspace*{.1in}}
}
\author{Shufang Su}
\affiliation{Department of Physics, University of Arizona, Tucson, AZ
85721, USA
\PRE{\vspace*{.5in}}
}

\author{Fumihiro Takayama%
\PRE{\vspace*{.2in}}
} 
\affiliation{Department of Physics and Astronomy, University of
California, Irvine, CA 92697, USA
\PRE{\vspace*{.1in}}
}

%\date{March 2004}

\begin{abstract}
\PRE{\vspace*{.3in}} Dark matter may be composed of superWIMPs,
superweakly-interacting massive particles produced in the late decays
of other particles.  We focus on the case of gravitinos produced in
the late decays of sleptons or sneutrinos and assume they are produced
in sufficient numbers to constitute all of non-baryonic dark matter.
At leading order, these late decays are two-body and the accompanying
energy is electromagnetic.  For natural weak-scale parameters, these
decays have been shown to satisfy bounds from Big Bang nucleosynthesis
and the cosmic microwave background.  However, sleptons and sneutrinos
may also decay to three-body final states, producing hadronic energy,
which is subject to even more stringent nucleosynthesis bounds.  We
determine the three-body branching fractions and the resulting
hadronic energy release.  We find that superWIMP gravitino dark matter
is viable and determine the gravitino and slepton/sneutrino masses
preferred by this solution to the dark matter problem.  In passing, we
note that hadronic constraints disfavor the possibility of superWIMPs
produced by neutralino decays unless the neutralino is photino-like.
\end{abstract}

%95.35.+d Dark matter
%98.80.Cq Particle-theory and field-theory models of the early Universe
%26.35.+c Big Bang nucleosynthesis
%98.80.Es Observational cosmology (including Hubble constant, 
%          distance scale, cosmological constant, early Universe, etc)

\maketitle

\section{Introduction}
\label{sec:introduction}

SuperWIMPs, superweakly interacting massive particles produced in the
late decays of weakly-interacting massive particles (WIMPs), are
promising non-baryonic dark matter
candidates~\cite{Feng:2003xh,Feng:2003uy}. Well-motivated superWIMP
candidates are the gravitino in supersymmetric
models~\cite{Feng:2003xh,Feng:2003uy,Ellis:1984er,Bolz:1998ek,%
Bolz:2000fu,Ellis:2003dn} and the first excited graviton in universal
extra dimension models~\cite{Feng:2003xh,Feng:2003nr}.

The supersymmetric possibility is realized naturally in supergravity
with a gravitino lightest supersymmetric particle (LSP) and a slepton
or sneutrino next-to-lightest supersymmetric particle
(NLSP). (Throughout this paper, ``slepton'' denotes a charged
slepton.) Both the gravitino and the NLSP have weak-scale masses.  The
NLSP freezes out as usual with a relic density near the observed
value.  However, after time $t \sim 10^4 - 10^8~\s$, it decays through
\begin{eqnarray}
\tilde{l} &\to& l \, \tilde{G} \nonumber \\
\tilde{\nu} &\to& \nu \, \tilde{G} \ .
\label{decay}
\end{eqnarray}
The gravitino then inherits much of the relic density of the slepton
or sneutrino~\cite{Covi:1999ty}, and its relic density is naturally of
the right magnitude without the introduction of new scales.  This is
in contrast to other gravitino dark matter scenarios, where the
gravitino is a thermal relic and the desired density is obtained by an
appropriately chosen gravitino mass $m_{\Gravitino} \sim
\kev$~\cite{Pagels:ke}, or the gravitino is produced during
reheating~\cite{Moroi:1993mb,Bolz:2000fu}, where the desired dark
matter density is obtained only for a fine-tuned reheat temperature
$T_R \sim 10^{10}~\gev$.

The decays of \eqref{decay} occur well after Big Bang nucleosynthesis
(BBN).  An immediate concern, therefore, is that they might destroy
the successful light element abundance predictions of BBN.  In fact,
BBN is not the only worry --- the Planckian spectrum of the cosmic
microwave background (CMB), the diffuse photon background, and bounds
on late time entropy production from the coincidence between BBN and
CMB baryometry all impose significant constraints.  As demonstrated in
Refs.~\cite{Feng:2003xh,Feng:2003uy}, however, these constraints on
the electromagnetic (EM) energy released in the decays of
\eqref{decay} exclude some of the weak scale parameter space, but
leave much of it intact.  Of course, at the border between the
excluded and allowed regions, slight deviations from standard
cosmological predictions are expected, providing possible signals in
future observations.  In fact, the existing anomaly in the $^7$Li
abundance prediction of standard BBN may already be interpreted as a
signal of superWIMP dark matter~\cite{Feng:2003uy}.  Such signals are
particularly welcome, since superWIMP dark matter is so weakly
interacting that it is impossible to detect through conventional
direct and indirect dark matter searches.

The previous work, however, neglected hadronic energy produced in WIMP
decays.  This was natural, as the WIMP decays of \eqref{decay}
contribute only to electromagnetic energy, as we will discuss below.
However, the three-body decays
\begin{eqnarray}
\tilde{l} &\to& l Z \tilde{G} \ , \ \nu W \tilde{G} \nonumber \\
\tilde{\nu} &\to& \nu Z \tilde{G} \ , \ l W \tilde{G} \ ,
\end{eqnarray}
produce hadronic energy when the $Z$ or $W$ decays hadronically.
Hadronic energy release is very severely constrained by the observed
primordial light element abundances~\cite{Reno:1987qw,%
Dimopoulos:1987fz,Dimopoulos:1988ue,Kohri:2001jx,Kawasaki:2004yh},
and so even subdominant hadronic decays could, in principle, provide
stringent constraints.  These three-body decays may be kinematically
suppressed when $m_{\tilde{l}, \tilde{\nu}} - m_{\tilde{G}} < m_W,
m_Z$, but even in this case, four-body decays, such as $\tilde{l} \to
l \gamma^* \tilde{G} \to l q\bar{q} \tilde{G}$, contribute to hadronic
cascades and may be important.  In this paper, we determine the
hadronic branching fractions and compare them to BBN constraints on
hadronic energy release.  Although we focus on the supersymmetric
case, our results may be extended straightforwardly with minor
numerical modifications to the case of graviton superWIMPs in extra
dimensions.

In evaluating the constraints, there are two possible approaches.  As
the superWIMP relic abundance is automatically in the right range, a
natural assumption is that superWIMP gravitinos make up all of the
non-baryonic dark matter, with
\begin{equation}
\OmegaSWIMP \simeq 0.23 \ .
\label{omega}
\end{equation}
This is the approach taken here.  Note that $\OmegaSWIMP$ need not be
identical to $\Omega_{\gravitino}$ --- not all relic gravitinos need
be produced by NLSP decays.  Our assumption therefore requires that
the other sources of gravitinos be insignificant.  For the
supersymmetric case, this typically implies an upper bound on reheat
temperatures of $T_R \alt 10^{10}~\gev$~\cite{Moroi:1993mb,Bolz:2000fu}.
Higher reheat temperatures may also be
allowed~\cite{Buchmuller:2003is,Kolb:2003ke} and are desirable, for
example, to accommodate leptogenesis.  For the universal extra
dimension case, the requirement of insignificant Kaluza-Klein graviton
production during reheating implies $T_R \alt 1$ to $10^2~\tev$,
depending, in part, on the number of extra
dimensions~\cite{Feng:2003nr}.

On the other hand, the Universe has proven to be remarkably baroque,
and there is no guarantee that dark matter is composed of only one
component.  One might therefore relax the constraint of \eqref{omega}
on the superWIMP energy density and assume, for example, that the NLSP
freezes out with its thermal relic density
$\Omega_{\text{NLSP}}^{\text{th}}$.  The superWIMP gravitino density
is then $\OmegaSWIMP = (m_{\gravitino}/m_{\text{NLSP}})
\Omega_{\text{NLSP}}^{\text{th}}$.  In this approach, the gravitino
density may be low and even insignficant cosmologically.  {}From a
particle physics viewpoint, however, the viability of the gravitino
LSP scenario is still worth investigation, as it has strong
implications for the superpartner spectrum and collider signatures,
independent of its cosmological importance.

These approaches differ significantly, not only in the viewpoint
taken, but also in their implications.  Suppose, for example, that the
gravitino and NLSP masses are both parametrized by a mass scale
$m_{\text{SUSY}}$.  The NLSP number density then scales as
$1/m_{\text{SUSY}}$ if one assumes \eqref{omega}, but scales as
$m_{\text{SUSY}}$ if one assumes a thermal relic NLSP density, since
$\Omega_{\text{NLSP}}^{\text{th}} \propto \langle \sigma v \rangle
^{-1} \propto m_{\text{SUSY}}^2$, where $\langle \sigma v \rangle$ is
the thermally-averaged NLSP annihilation cross section.  Low masses
are excluded in the former case, while high masses are disfavored in
the latter.  We consider both approaches to be worthwhile, but defer
discussion of the thermal relic density approach, along with its very
different consequences for the superpartner spectrum and implications
for collider physics, to a separate study~\cite{Feng:2004}.

Assuming a fixed $\OmegaSWIMP$ here, we find that gravitino superWIMPs
provide a viable solution to the dark matter problem.  We determine
the allowed masses for the NLSP and gravitino.  At the same time, we
find that the hadronic energy constraint is significant and does in
fact provide the leading constraint in natural regions of parameter
space. We conclude that the analysis below must be done in any
scenario with similar late decays.  Given any standard cosmology, a
significant thermal relic abundance of NLSPs will be generated, these
NLSPs will eventually decay to superWIMPs, and the hadronic
constraints we discuss below must be analyzed before the scenario may
be considered viable.

In passing, we note that another possibility is that the NLSP is the
lightest neutralino.  For a general neutralino, the decays $\chi \to Z
\tilde{G}, h \tilde{G}$ produce large hadronic energy release.  These
two-body modes are absent for photino-like neutralinos, where the only
two-body decay is $\tilde{\gamma}\to\gamma \tilde{G}$. However, even
in this case, one cannot avoid hadronic activity from decay through a
virtual photon, $\tilde{\gamma} \to \gamma^* \tilde{G} \to q\bar{q}
\tilde{G}$, with a branching fraction of ${\cal O} (10^{-2})$. Given
the stringency of constraints on hadronic energy release, the
neutralino NLSP case is highly constrained.  Assuming that
$\OmegaSWIMP$ makes up a significant fraction of the dark matter,
natural regions of parameter space are excluded~\cite{Feng:2004}.  In
the current paper, we therefore focus on the more promising
possibilities of slepton and sneutrino NLSPs.

The paper is organized as follows.  In Sec.~\ref{sec:bbn}, we
summarize what is known about constraints on EM and hadronic energy
release in the early Universe.  In Sec.~\ref{sec:swimp}, we give a
detailed description of the gravitino superWIMP scenario with a
slepton or sneutrino NLSP.  In Sec.~\ref{sec:had}, we determine the
hadronic energy release from slepton and sneutrino decays and find the
allowed mass regions.  We summarize in Sec.~\ref{sec:conclusion}.  The
relevant gravitino interactions, decay amplitudes, and three-body
hadronic decay width formulae are collected in the Appendix.

\section{Late time energy injection and BBN}
\label{sec:bbn}

The overall success of standard BBN places severe constraints on
energy produced by particles decaying after BBN.  Given the baryon
density specified by CMB measurements~\cite{Spergel:2003cb}, the
prediction of standard BBN~\cite{Burles:2000zk} for deuterium agrees
well with observations~\cite{Kirkman:2003uv}.  This concordance is
less perfect, but still reasonable, for
$^4$He~\cite{Olive:1996zu,Izotov}.  On the other hand, current
observations of $^7$Li~\cite{Thorburn,Bonafacio,Ryan:1999vr} are
consistently lower than predictions; this is the leading anomaly in
light element abundances at present.  The abundances of $^3$He and
$^6$Li also provide constraints, although these are subject to more
theoretical uncertainties as discussed below.

\begin{figure}[tbp]
\postscript{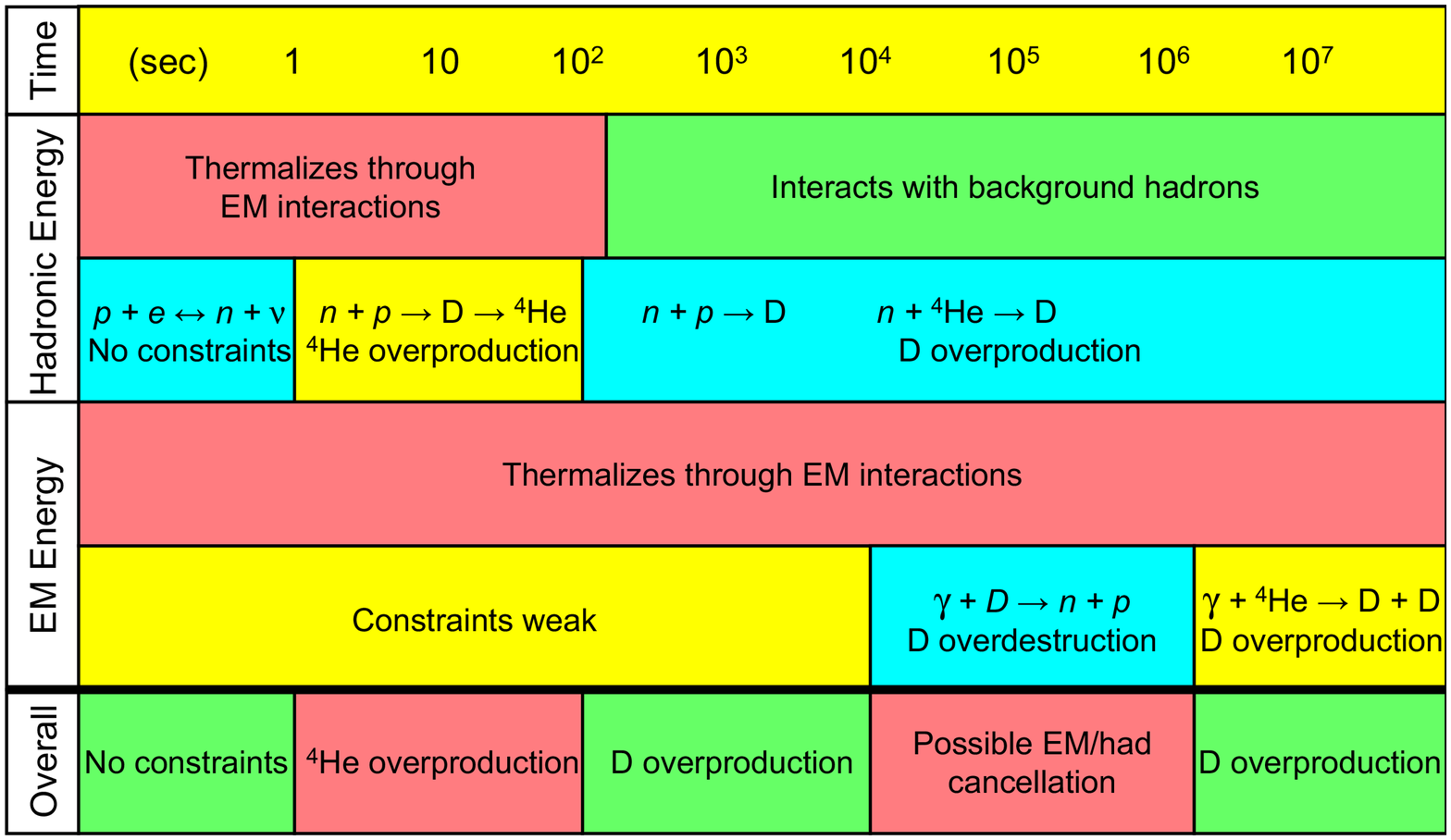}{0.9}
\caption{Summary of the leading constraints from D and $^4$He on late
time EM and hadronic energy injection into the early Universe.
Constraints from $^3$He and $^6$Li may also be important.  (See
text.) }
\label{fig:bbnconstraints}
\end{figure}

In this section, we review the constraints from these element
abundances on energy release in the early Universe.  Although a
unified analysis is necessary for detailed conclusions, we review the
constraints on EM and hadronic energy release in turn.  A schematic
summary of the discussion in this section is given in
Fig.~\ref{fig:bbnconstraints}.

The constraints on EM energy release are well understood, and we begin
with a brief summary of the results.  (For details, see
Ref.~\cite{Cyburt:2002uv}.)  High energy photons and leptons produced
by late-decaying particles are quickly thermalized through
interactions with background (BG) particles.  High energy photons, for
example, initiate EM cascades through processes like
$\gamma\gamma_{\text{BG}}\to e^+e^-$ and $\gamma e_{\text{BG}}\to
\gamma e$, and the initial energy is quickly converted to very soft
photons.  The same is true for high energy electrons and muons.  High
energy taus decay to mesons, but these decay and produce EM cascades
before interacting hadronically. (See Ref.~\cite{Feng:2003uy} and the
analysis below.)  Injected leptons of all three families therefore
contribute dominantly to EM energy and negligibly to hadronic energy.

The energy distribution of the resulting photons is highly suppressed
for $E_{\gamma} > E_{\text{max}}$, where
\begin{eqnarray}
E_{\text{max}} = \frac{m_e^2}{22\, T} \simeq
12~\mev \left[ \frac{\kev}{T} \right]
\end{eqnarray} 
is related to the energy threshold for $\gamma\gamma_{\text{BG}}\to
e^+e^-$~\cite{Kawasaki:1994sc}.  Here $T$ is the BG photon
temperature; in the radiation-dominated era, it is related to time by
\begin{eqnarray}
 t\simeq 1.0\times 10^6~\s \left[\frac{{\kev}}{T}\right]^2 \ .
\end{eqnarray}
Because of this rapid thermalization, constraints on EM energy depend
on the total amount of energy released and are essentially independent
of the shape of the initial energy spectrum.

For $t \alt 10^4~\s$, the BBN constraints are rather weak since
$E_{\text{max}}$ is too small to destroy any light elements.  For
$10^4~\s \alt t \alt 10^6~\s$, the Universe is hot and
$E_{\text{max}}$ is low, and so only light elements with very low
binding energies can be destroyed.  The main constraint is from the
overdestruction of D, with a binding energy of only 2.2 MeV.  For $t
\agt 10^7~\s$, the Universe is cooler and $E_{\text{max}}$ becomes
high enough that $^4$He, with binding energy 19.8 MeV, is destroyed.
Even a small amount of $^4$He photo-dissociation becomes a significant
production mechanism for $^3$He and D, since the observed abundances
of $^3$He and D are much lower than that of $^4$He.  The main
constraint is from the overproduction of D.

Between $10^6~\s$ and $10^7~\s$ there is a region where the
overdestruction and overproduction of D cancel.  For sufficiently
high energy release in this region, the D abundance is that of
standard BBN, but $^7$Li, with binding energy 2.5 MeV, is destroyed
through $^7\text{Li}+\gamma \to n+ ^6\text{Li}$.  The observed $^7$Li
abundance~\cite{Thorburn,Bonafacio,Ryan:1999vr} is in fact low
compared to the predictions of standard BBN~\cite{Burles:2000zk}. We
see, then, that the $^7$Li abundance can be brought into agreement
with observations without upsetting the D concordance for decay times
in this window. For $\eta=n_B/n_{\gamma}=6.0\times 10^{-10}$, the best
fit values are~\cite{Cyburt:2002uv}
\begin{eqnarray}
\xi_{\text{EM}} \approx 10^{-9}~\gev\ , \quad
t \approx 3 \times 10^6~\s \ ,
\label{bestfit}
\end{eqnarray}
where 
\begin{eqnarray}
\xi_{\text{EM}} \equiv \epsilon_{\text{EM}} B_{\text{EM}} Y_{\text{NLSP}} 
\label{eq:xi_EM}
\end{eqnarray}
is the emitted EM energy density. Here $\epsilon_{\text{EM}}$ is the
initial EM energy released in NLSP decay, and $B_{\text{EM}}$ is the
branching fraction of NLSP decay into EM components.  $Y_{\text{NLSP}}
\equiv n_{\text{NLSP}}/n_{\gamma}$ is the NLSP number density just
before NLSP decay, normalized to the BG photon number density
$n_{\gamma} = 2 \zeta(3) T^3 / \pi^2$.  It can be expressed in terms
of the superWIMP abundance:
\begin{equation}
Y_{\text{NLSP}}\simeq 3.0 \times 10^{-12}
\left[\frac{\tev}{m_{\text{SWIMP}}}\right]
\left[\frac{\OmegaSWIMP}{0.23}\right] \ .
\label{eq:def_Y}
\end{equation}
The parameters of \eqref{bestfit} are naturally realized in the
superWIMP scenario when a weak-scale NLSP decays to a weak-scale
superWIMP with the observed dark matter density~\cite{Feng:2003uy}.

The numerical constraints on EM energy release as a function of the
particle decay lifetime have been determined in
Ref.~\cite{Cyburt:2002uv}, using the following criteria:
\begin{eqnarray}
\text{D\ low}:&&\text{D/H}<1.3\times 10^{-5} \label{D} \\
\text{D\ high}:&&\text{D/H}>5.3\times 10^{-5} \\
^4\text{He\ low}:&&Y_p<0.227 \\
^7\text{Li\ low}:&&^7\text{Li/H}<0.9\times 10^{-10} \ .
\label{EMconstraints}
\end{eqnarray}
We adopt these constraints in our consideration of EM energy release
bounds.

Hadronic energy release, and its effects on BBN, has also been
considered~\cite{Reno:1987qw,Dimopoulos:1987fz,Dimopoulos:1988ue,%
Kohri:2001jx,Kawasaki:2004yh}.  Although we are most interested in
energy release at times $t \agt 10^4~\s$, we summarize here what is
known for $t \agt 1~\s$ for completeness.

As we will show below, for the decay times of interest, mesons decay
quickly and so contribute to EM energy.  Only protons and neutrons are
sufficiently long-lived that they might be able to interact
hadronically.  For times $t \alt 150~\s$, however, even high energy
protons and neutrons (with energies $\agt 100~\gev$) interact through
EM scattering with BG photons before interacting with BG baryons.
Injected high energy baryons then quickly become slow baryons, with
the initial energy dissipated to EM energy.  For such early times,
this EM energy is harmless, as discussed above.  Constraints are
therefore functions of the number density of injected baryons
$B_{\text{had}} Y_{\text{NLSP}}$, and independent of the initial
baryon energy $\epsilon_{\text{had}}$.

Slow baryons modify the number of neutrons and protons and therefore
BBN predictions.  In the very earliest era with $t \alt 1~\s$,
however, the weak interaction $p+e\leftrightarrow n+\nu$ is efficient,
and so any deviation in the $n/p$ ratio is washed out, and there are
no strong bounds.  For $t \agt 1~\s$, the weak interactions decouple.
Slow baryons produced in late decays cause the $n/p$ ratio to deviate
from the standard BBN value.  In particular, extra neutrons create D.
For $1~\s \alt t \alt 100~\s$, the resulting D is immediately burned
to form $^4$He, and so the dominant constraint is from overproduction
of $^4$He.  This constraint is weak since the $^4$He abundance of
standard BBN is already large.  For $t>100~\s$, D can no longer burn
into $^4$He~\cite{Reno:1987qw, Kohri:2001jx} and the dominant
constraint comes from D overproduction.  This bound is significantly
stronger, since the D/H ratio is $\sim 10^{-5}$.

At times $t \agt 150~\s$, injected baryons interact with BG baryons
before thermalizing.  These interactions create secondary hadrons that
can modify BBN predictions.  In this era, constraints depend in
principle on both $\epsilon_{\text{had}}$ and $B_{\text{had}}
Y_{\text{NLSP}}$ separately.  In principle, the constraints do not
simply depend on the product, because the total number of secondary
hadrons created by an injected hadron is not always proportional to
the injected hadronic energy.  The dominant constraint is from
secondary baryons destroying $^4$He to overproduce D.

Quantitative analyses on the BBN constraints of the hadronic energy
injection have been performed in Refs.~\cite{Reno:1987qw,%
Dimopoulos:1987fz,Dimopoulos:1988ue,Kohri:2001jx,Kawasaki:2004yh}.  As
discussed above, formally, the constraints depend on the lifetime
$\tau$ of the decaying particle and on $\epsilon_{\text{had}}$ and
$B_{\text{had}} Y_{\text{NLSP}}$ separately.  However, the most recent
analysis~\cite{Kawasaki:2004yh} finds that for decay times $\tau \alt
10^6~\s$, when the hadronic constraint is most important, the
constraints depend on the product $\epsilon_{\text{had}}
B_{\text{had}} Y_{\text{NLSP}}$ to within a factor of 2 for $100~\gev
\le \epsilon_{\text{had}} \le 1~\tev$~\cite{Kohri}.  We will therefore
assume this behavior and apply the hadronic constraint to times $\tau
\alt 10^6~\s$, where it is stronger than the EM constraint in the
superWIMP scenario.

The authors of Ref.~\cite{Kawasaki:2004yh} assume the following
constraints on the light element abundances (95\% CL):
\begin{eqnarray}
&&\text{D}/\text{H}=(2.8 \pm 0.8) \times 10^{-5} 
\label{Dconstraints}\\
&& Y_{{p}}=0.238 \pm 0.004 \pm 0.010 
%\ (\text{FO}) 
\\
&& \text{log}_{10}({}^7\text{Li}/\text{H})=-9.66 \pm 0.112 \pm 0.6 \ .
\label{hadconstraints}
\end{eqnarray}
In our analysis of hadronic energy injection, we take the constraints
of Ref.~\cite{Kawasaki:2004yh} derived using these bounds and
$B_{\text{had}}=1$.

Note that the constraints of
Eqs.~(\ref{Dconstraints})-(\ref{hadconstraints}) adopted for the
hadronic energy analysis are less conservative than the constraints of
Eqs.~(\ref{D})-(\ref{EMconstraints}) used in the EM analysis of
Ref.~\cite{Cyburt:2002uv}.  In particular, a significantly more
aggressive D bound is assumed in the hadronic analysis.  There is now
impressive concordance between the baryon number determinations from D
and CMB measurements, which further supports the narrow range of D/H
given in \eqref{Dconstraints}.  At the same time, existing anomalies
in standard BBN may be indications that systematic errors are
underestimated or that there is new physics, which would likely
distort the D abundance, since D is very weakly bound.

There are several further comments to make.  First, we have been
treating the EM and hadronic effects separately.  Of course, these
effects can add constructively or
destructively~\cite{Dimopoulos:1988ue}.  We have explained above that
the dominant hadronic constraint for $t \agt 10^4~\s$ is from D
overproduction, whereas for the EM constraint it is from D
overproduction for $t \agt 10^6~\s$ and from D overdestruction for
$10^4~\s \alt t \alt 10^6~\s$.  For phenomena contributing to both EM
and hadronic energy, then, the effects add constructively for $t \agt
10^6~\s$, but may cancel for $10^4~\s \alt t \alt 10^6~\s$, and there
may be fingers of allowed region in the latter time interval.

Finally, although we consider only the constraints above on D/H, $Y_p$
and ${^7}$Li/H, in Ref.~\cite{Kawasaki:2004yh}, constraints from
$^3$He/D and $^6$Li/H are also analyzed for late time EM and hadronic
injections.  These two abundances, however, are subject to more
uncertainties.  $^3$He/D is considered because the destruction of
$^4$He could produce $^3$He more easily than D, while the binding
energy of $^3$He is higher than that of D.  However, the
interpretation of $^3$He measurements~\cite{Bania} may be subject to
ambiguities arising from stellar production and
destruction~\cite{Vangioni-Flam:2002sa}.  The ratio of $^6$Li/$^7$Li
has been observed in low metallicity halo stars and might also
constrain late decays.  The implications for EM cascades have also
been explored in Refs.~\cite{Brown:1988pb,Jedamzik:1999di}.  Again,
however, there are arguably significant systematic uncertainties
arising from the higher depletion of $^6$Li relative to $^7$Li as a
result of its smaller binding energy~\cite{Nissen:1999iq}.  A better
understanding of the connection between the primordial and observed
abundances of $^6$Li and $^7$Li could provide a very powerful
constraint on late decaying particles~\cite{Jedamzik:2004er}.

\section{Slepton and sneutrino NLSP decays}
\label{sec:swimp}

In the superWIMP dark matter scenario, superWIMPs have no effect on
the early thermal history of the Universe.  They are produced only
through metastable NLSP decays, NLSP$\to\tilde{G}$ + SM particle,
giving a gravitino relic density of
\begin{eqnarray}
\OmegaSWIMP = \frac{m_{\tilde{G}}}
{m_{\text{NLSP}}} \Omega_{\text{NLSP}}^{\text{th}} \ .
\end{eqnarray}
$\Omega_{\text{NLSP}}^{\text{th}} > 0.23$ is allowed, since
$m_{\tilde{G}}/m_{\text{NLSP}}<1$.  The SM particles produced in the
NLSP decay induce EM or hadronic cascades, which will affect the
primordial abundance of light elements, as discussed in
Sec.~\ref{sec:bbn}.

The dominant decay mode for slepton NLSPs is $\tilde{l}\to l
\tilde{G}$.  Most of the energy released from the decay is transferred
to EM cascades.  The total energy released in slepton decay is
\begin{eqnarray}
E_{\text{total}} =
\frac{m_{\text{NLSP}}^2-m_{\tilde{G}}^2}{2m_{\text{NLSP}}} \ .
\end{eqnarray}
The total injected EM energy density is, for $\Delta m = m_{\tilde{l}}
- m_{\tilde{G}} \alt m_{\tilde{l}}$,
\begin{eqnarray}
\xi_{\text{EM}} \equiv \epsilon_{\text{EM}}B_{\text{EM}}
Y_{\tilde{l}}\sim {\cal O}(10^{-9}\  \text{GeV})
\left[\frac{\epsilon_{\text{EM}}B_{\text{EM}}}{E_{\text{total}}}\right]
\left[\frac{\Delta m}{300~\gev}\right]
\left[\frac{\tev}{m_{\tilde{G}}}\right]
\left[\frac{\OmegaSWIMP}{0.23}\right] \ ,
\end{eqnarray}
where $\epsilon_{\text{EM}} B_{\text{EM}}$ depends on the slepton's
flavor, and Eq.~(\ref{eq:def_Y}) has been used.  For selectrons and
smuons, energetic electrons and muons quickly thermalize with the
background photons.  All the visible energy is transferred to the
resulting EM cascade: $\epsilon_{\text{EM}}=E_{\text{total}}$,
$B_{\text{EM}}=1$.  In the case of staus, the decays of taus could
also produce mesons and neutrinos, in addition to electrons and muons.
The EM activity depends on the lifetime and energy distribution of
decay products.  As estimated in Ref.~\cite{Feng:2003uy}, the EM
energy released in stau decay is in the range $0.3 E_{\text{total}}$
to $E_{\text{total}}$.

For the sneutrino NLSP case, the sneutrino mainly decays into neutrino
and gravitino. The emitted high energy neutrino interacts with BG
neutrinos and loses energy.  If the neutrinos annihilate through $\nu
\nu_{\text BG}\to e^+e^-$, an EM cascade is initiated.  However, the
annihilation takes place through weak interactions, and so its cross
section may be small.  The effect of the expansion of the Universe
must also be taken into account~\cite{Kawasaki:1994bs,
Frieman:1989vt}.  Sneutrino decays induce far less EM energy release
than sleptons, but for decay times $\tau \alt 10^7~\s$, there may
still be some impact on BBN.  For such decays times, however, much of
the parameter space is excluded by considerations of hadronic BBN
constraints, which will be discussed below.  In what is left of this
region, the gravitino mass is $m_{\tilde{G}} < {\cal O}(10)~\gev$.
Such light gravitinos are less motivated in supergravity and so we
have not considered this scenario in great detail here.

\begin{figure}
\resizebox{6.5 in}{!}{
\includegraphics{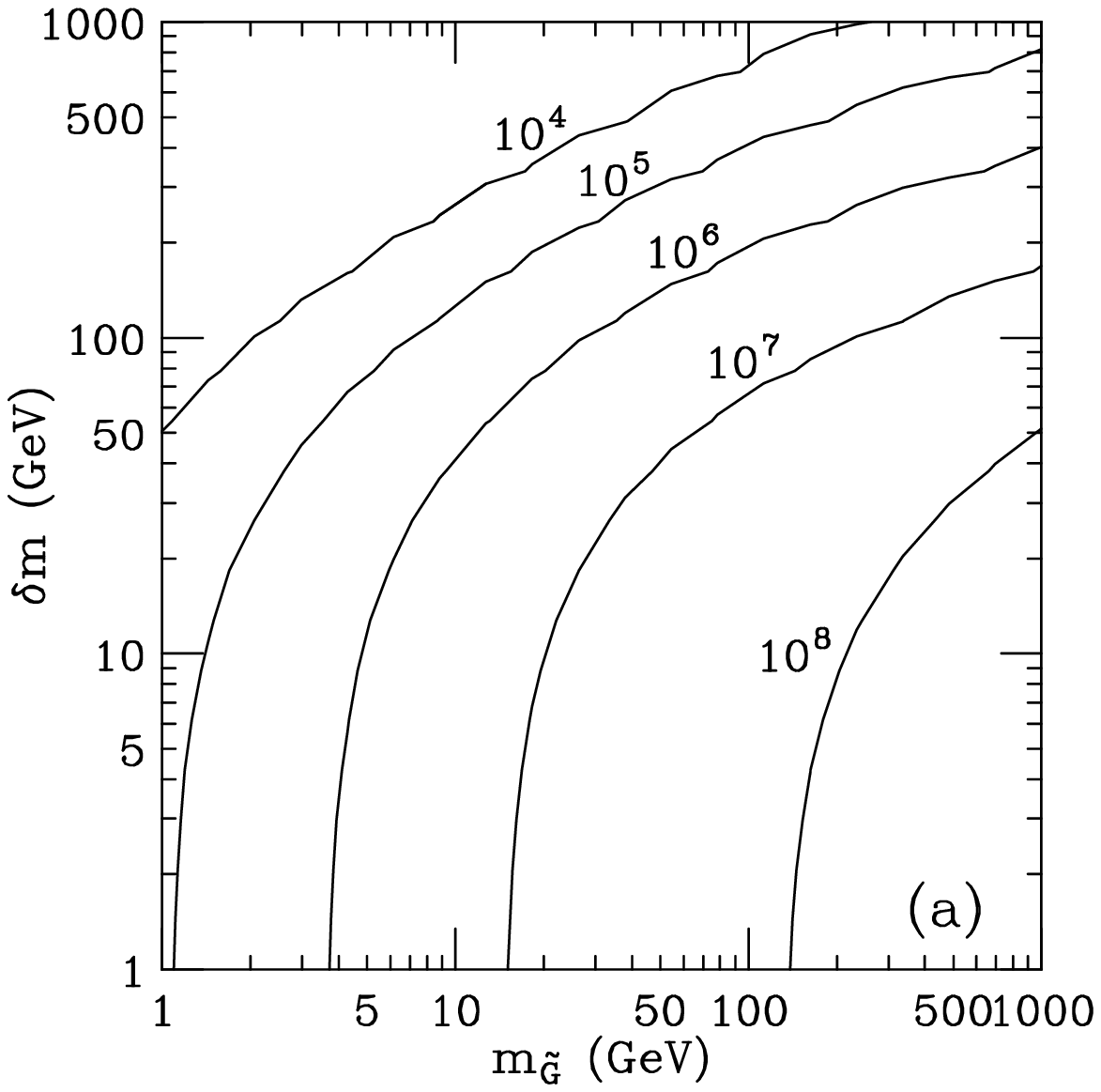}
\includegraphics{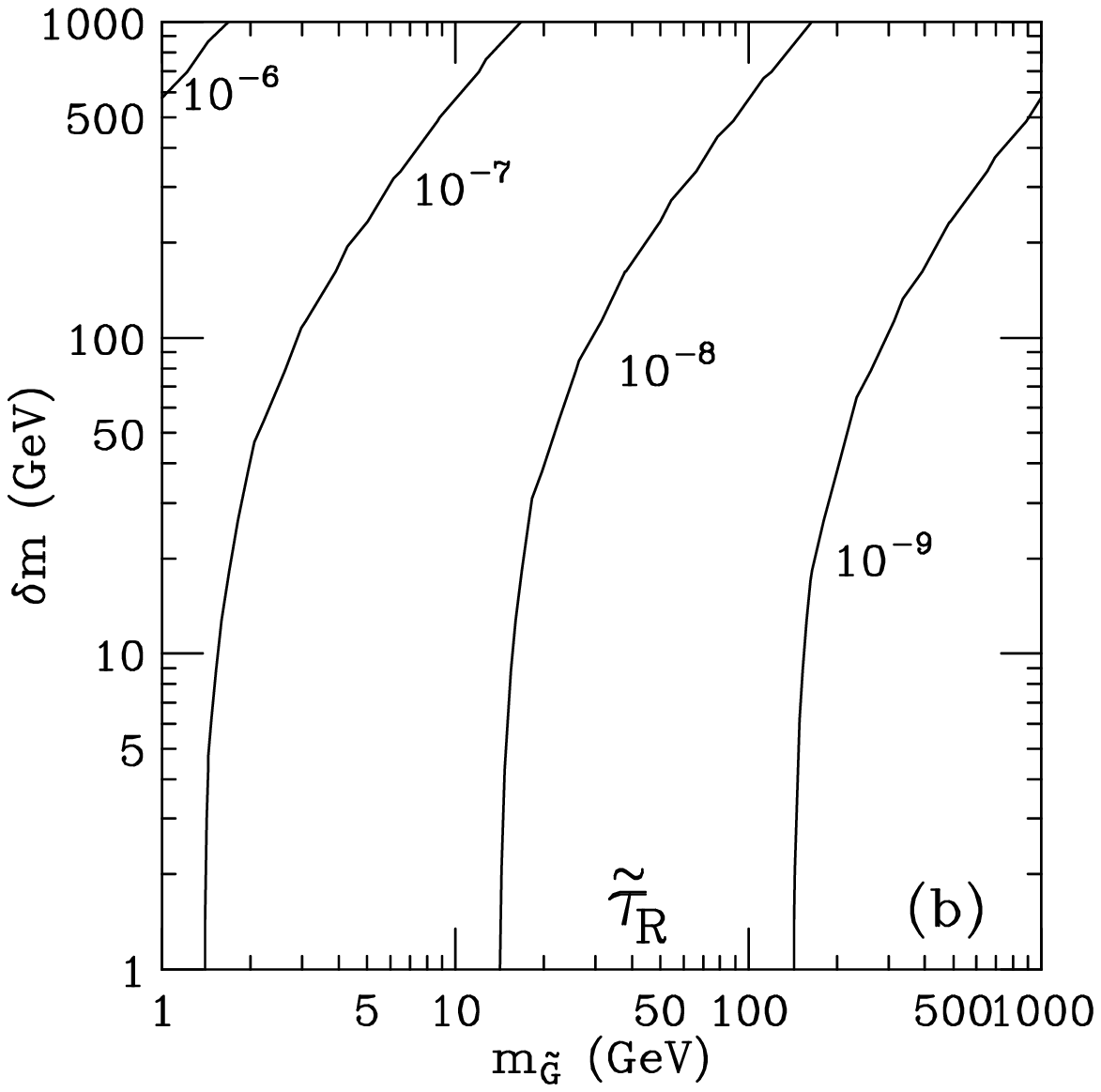}}
\caption{(a) Lifetime (in seconds) for $\tilde{l}_{L,R}$,
$\tilde{\nu}$ NLSPs and (b) EM energy release (in GeV) for the
$\tilde{\tau}_R$ NLSP as functions of gravitino mass $m_{\tilde{G}}$
and mass difference $\delta m = m_{\tilde{l}} - m_{\tilde{G}} -
m_Z$.  For (b), we take $\epsilon_{\text EM}=0.5 E_{\text{total}}$,
$B_{\text{EM}}=1$, and $\OmegaSWIMP = 0.23$.  }
\label{fig:lifetime_zetaEM}
\end{figure}

The decay width of the slepton into LSP gravitino is
\begin{eqnarray}
 \Gamma(\tilde{l}_{L,R}\to l_{L,R}\tilde{G}) =\frac{1}{96\pi M_4^2}
 \frac{m_{\tilde{l}}^5}{m_{\tilde{G}}^2}\left[1
-\frac{m_{\tilde{G}}^2}{m_{\tilde{l}}^2}
 \right]^4 \ ,
\label{eq:sleptondecay}
\end{eqnarray}
where $M_4=(16\pi G_N)^{-1/2}$.  For $\mgravitino / m_{\tilde{l}}
\approx 1$, the slepton decay lifetime is
\begin{eqnarray}
 \tau(\tilde{l}_{L,R}\to l_{L,R} \tilde{G})
\simeq 3.6\times 10^8~\s
\left[\frac{100~\gev}{\Delta m}\right]^4
\left[\frac{m_{\tilde{G}}}{\tev}\right]\ .
\label{eq:decaylifetime}
\end{eqnarray}
This expression is valid only when the gravitino and slepton are
nearly degenerate, but we present it here as a useful guide for the
reader.  In the analysis below, we always use \eqref{eq:sleptondecay}
for the evaluation of the slepton decay lifetime.  The slepton
lifetime is given in Fig.~\ref{fig:lifetime_zetaEM}a in the
$(m_{\tilde{G}},\delta m)$ plane, where
\begin{equation}
\delta m \equiv \Delta m -m_Z
= m_{\tilde{l}} - m_{\tilde{G}} - m_Z \ .
\end{equation}
Note that in the definition of $\delta m$, an additional $m_Z$ is
subtracted from $\Delta m$; three-body decays, to be discussed below,
are therefore always kinematically possible for $\delta m > 0$.  The
typical decay time is $10^4~\s \alt \tau \alt 10^8~\s$.  For such
decay times, there are strong constraints from the primordial
abundance of the light element, as discussed earlier.

Contours of EM energy release $\xi_{\text{EM}}=\epsilon_{\text
EM}B_{\text EM}Y_{\tilde{l}}$ are given in
Fig.~\ref{fig:lifetime_zetaEM}b for $\tilde{\tau}_R$ NLSPs, assuming
$\epsilon_{\text{EM}}=0.5 E_{\text{total}}$, $B_{\text{EM}}=1$, and
$\OmegaSWIMP = 0.23$.  We see that the EM energy release varies from
$10^{-6}-10^{-9}~\gev$ and is largest for large $\delta m$ and small
$m_{\tilde{G}}$, when the decay lifetime is the shortest.  The BBN
constraints, however, are much weaker for early decays.  As discussed
in Ref.~\cite{Feng:2003uy}, considering EM energy constraints only,
one finds that for a stau NLSP, $m_{\tilde{G}} \sim 300-500~\gev$ and
$\delta m \sim 200-1000~\gev$ are allowed.  Moreover, late time EM
energy release from stau decays to gravitinos may resolve the current
discrepancy in the $^7$Li for $(m_{\tilde{G}}, \delta m) \approx
(450~\gev, 200~\gev)$.  For selectron or smuon NLSPs, the EM
constraints are slightly stronger since all the released decay energy
is transferred to EM cascades.

The EM BBN constraints may be weakened by a number of possibilities.
First, for decays in the range $10^4~\s \alt \tau \alt 10^6~\s$, the
EM constraints may be weakened by cancelation between the effects of
EM and hadronic energy release, as discussed above.  Second, if not
all of non-baryonic dark matter is superWIMPs, $\OmegaSWIMP < 0.23$,
the contour lines for $\xi_{\text{EM}}$ will shift toward the left,
since $\xi_{\text{EM}}\propto \OmegaSWIMP$, and the allowed parameter
space will be enlarged.  Finally, if the NLSP is not a slepton but a
sneutrino, the EM BBN constraints are, of course, much weaker, as the
two-body decays contribute much less visible energy than in the
slepton NLSP case.

\section{Constraints from hadronic energy release}
\label{sec:had}

As discussed in Sec.~\ref{sec:bbn}, the abundances of light elements
constrain the amount of hadronic energy injection into the early
Universe.  Although, for leptonic NLSPs, the hadronic energy release
is much smaller than the EM energy release, the constraints on the
hadronic energy release are much stronger.  It is therefore important
to evaluate in detail the hadronic activity induced by slepton and
sneutrino NLSP decays.

Let us first study the slepton NLSP possibility.  The main decay mode
of sleptons is $\tilde{l}\to l \tilde{G}$.  For selectron and smuon
NLSPs, such decays induce only EM cascades.  For stau NLSPs, however,
the resulting tau lepton may decay into mesons, and these might
interact with background hadrons and induce dangerous hadronic
cascades.  The interaction time of the produced mesons is
\begin{eqnarray}
\tau_{\text{strong-int}} = 
\left[ \langle \sigma_{\text{had}} v \rangle n_B \right]^{-1}
&\simeq& 18~\s \left[\frac{100~\mb}
{\langle \sigma_{\text{had}}v \rangle} \right]
\left[\frac{6\times 10^{-10}}{\eta} \right]
\left[\frac{\kev}{T} \right]^3 \nonumber \\
&\simeq& 1.8\times 10^{-8}~\s 
\left[\frac{100~\mb}{\langle \sigma_{\text{had}} v \rangle} \right] 
\left[\frac{6\times 10^{-10}}{\eta} \right]
\left[\frac{t}{1~\s} \right]^{3/2} \ ,
\label{eq:interaction}
\end{eqnarray}
where $\langle \sigma_{\text{had}} v \rangle $ is the thermally
averaged strong interaction cross section.  For the decay times of
greatest interest here, $10^4~\s \alt \tau \alt 10^8~\s$, this
interaction time is long compared to the typical meson ($\pi$,$K$)
lifetime, $(E_{\text{Meson}}/m_{\text{Meson}})\times 10^{-8}~\s$.  The
produced mesons thus decay before they interact with background
hadrons.

\begin{figure}
\resizebox{6.5 in}{!}{\includegraphics*[0,630][600,730]{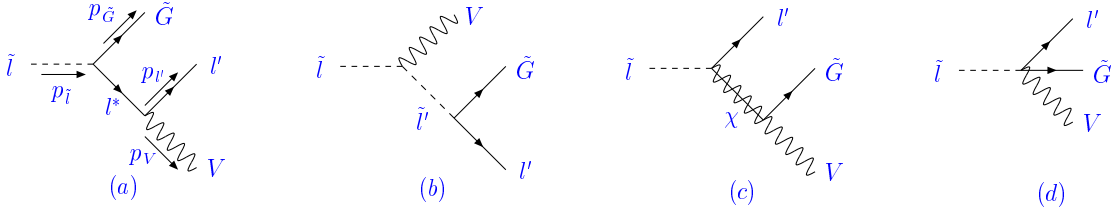}}
\caption{Feynman diagrams for slepton decays $\tilde{l}\to l' V
\tilde{G}$ leading to hadronic energy.  $\chi$ in diagram (c) is a
neutralino or chargino.  When kinematically accessible, $V$ is a $Z$
or $W$ boson.  For small $\Delta m < m_Z, m_W$, the leading hadronic
decay is $\tilde{l}\to l \gamma^* \tilde{G}$, followed by $\gamma^*
\to q \bar{q}$. }
\label{fig:feynmangraphs}
\end{figure}

As a result, the hadronic activity from the dominant two-body decay of
slepton is negligibly small.  The main contributions to hadronic
activity therefore come from three-body decays, $\tilde{l}\to l Z
\tilde{G}, \nu_l W \tilde{G}$ with $Z$ or $W$ decaying hadronically,
or from four-body decays $\tilde{l}\to l \gamma^* \tilde{G}$ with
$\gamma^*\to q\bar{q}$.  The corresponding Feynman diagrams are shown
in Fig.~\ref{fig:feynmangraphs}.  Formulae for the decay amplitudes
and decay width calculations are listed in the Appendix. 

The hadronic branching fraction is defined as follow:
\begin{equation}
B_{\text{had}} \equiv
\frac{\Gamma(\tilde{l}\to l Z \tilde{G}) B_{\text{h}}^Z
+\Gamma(\tilde{l}\to l' W \tilde{G})B_{\text{h}}^W
+\Gamma(\tilde{l}\to l' q \bar{q} \tilde{G})}
{\Gamma(\tilde{l}\to l \tilde{G})} \ ,
\label{hadfraction}
\end{equation}
where $B_{\text{h}}^Z , B_{\text{h}}^W \simeq 0.7$ are the $Z$ and $W$
hadronic branching fractions.  For right-handed sleptons,
$\Gamma(\tilde{l}\to \nu W \tilde{G}) = 0$, and there are additional
suppressions because the $Z$ couplings are smaller than the $W$
couplings.  Thus, the hadronic branching fraction of right-handed
sleptons is smaller than that of left-handed ones, especially when
$m_W <\Delta m<m_Z$.

For small $\Delta m$, the slepton lifetime increases dramatically,
since it grows as $(\Delta m)^{-4}$ by
Eq.~(\ref{eq:decaylifetime}). For $\Delta m <m_W$, three-body decay
modes are kinematically suppressed and four-body decay through a
virtual photon dominates.  The branching fraction of four-body decay
is roughly expected to be ${\cal O}((\alpha/4\pi)^2) \sim O(10^{-6})$,
which sets a lower limit for the hadronic branching fraction.
Adjusting this estimation by taking into account details of the
kinematics has negligible impact to our results.

For the sneutrino NLSP case, the constraints from bounds on EM energy
release are weak, especially at late times $\tau \agt 10^7~\s$, as
discussed in Sec.~\ref{sec:swimp}.  However, the hadronic energy
release is of the same magnitude for sneutrino NLSPs and for
left-handed slepton NLSPs.  The major hadronic activity comes from
three-body decays, $\tilde{\nu}\to \nu Z \tilde{G}, l W \tilde{G}$.
The branching fraction for the four-body decay $\tilde{\nu}\to \nu
\gamma^* \tilde{G} \to \nu q \bar{q} \tilde{G}$ is also similar to the
analogous decay in the slepton case, except that diagrams (a) and (b)
in Fig.~\ref{fig:feynmangraphs} are absent since the neutrino is
neutral.  Last, we note that in the special case where the Bino and
Wino masses are nearly equal, diagram (c) of
Fig.~\ref{fig:feynmangraphs} is highly suppressed, since the gauge
boson $V$ is almost purely $Z$.

\begin{figure}
\resizebox{6.5 in}{!}{
\includegraphics{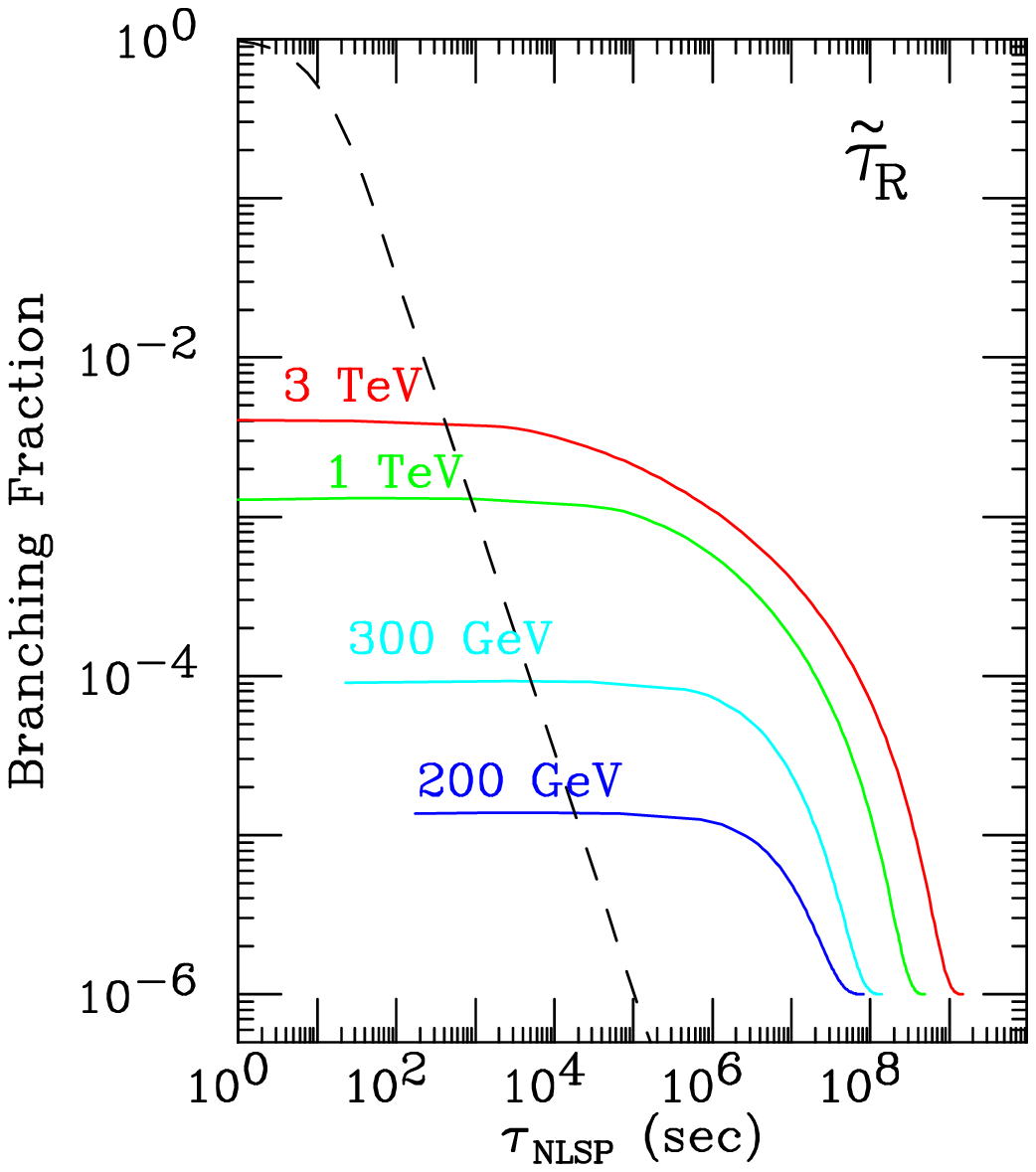}
\includegraphics{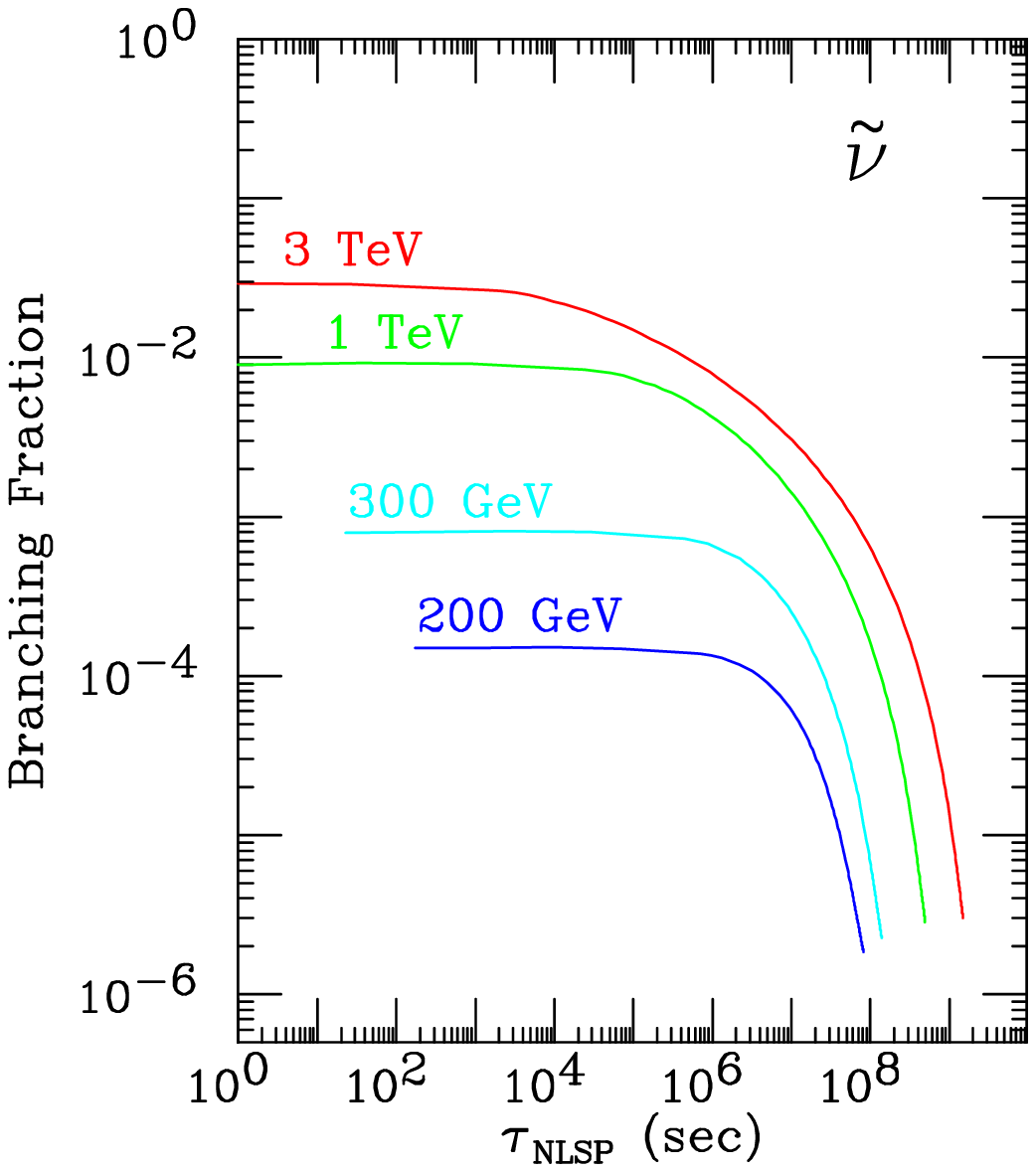}}
\caption{Hadronic branching fractions for (left) $\tilde{\tau}_R$
NLSPs and (right) $\tilde{\nu}$ NLSPs as functions of the NLSP decay
lifetime.  Along each curve, the NLSP mass is fixed at the value
indicated and the gravitino mass varies.  The neutralino/chargino
parameters are chosen to be $M_1=2 m_{\text{NLSP}}$, $M_2=\mu=4
m_{\text{NLSP}}$, and $\tan\beta=10$.  For the $\tilde{\tau}_R$ case,
the dashed line gives the hadronic branching fraction resulting from
$\tilde{\tau}_R \to \tau \gravitino$, where the tau decays to a meson,
and the meson interacts hadronically before decaying. We take the
meson rest lifetime to be $3\times 10^{-8}~\s$.}
\label{fig:branch}
\end{figure}

In Fig.~\ref{fig:branch} we show the hadronic branching fractions for
$\tilde{\tau}_R$ and $\tilde{\nu}$ NLSPs for a particular set of
values of $M_1$, $M_2$, $\mu$ and $\tan\beta$.  Varying these
parameters typically does not change our results significantly.  The
reason is that these parameters only affect the neutralino/chargino
appearing in diagram (c) of Fig.~\ref{fig:feynmangraphs}.  The
contribution from diagram (c) is negligible if the neutralino/chargino
is not degenerate to within about 10\% with the sleptons.  In
addition, diagram (c) does not interfere with the other diagrams if
all couplings are real, which eliminates the possibility of
cancelation.  Hadronic branching fractions for left-handed sleptons
are similar to those shown for sneutrinos.  The branching fractions
are smaller for the right-handed slepton case for the reasons
mentioned above.  In both cases, the hadronic branching fractions drop
quickly for lifetime $\tau \agt 10^6$ s, which corresponds to small
$\delta m$.  For the $\tilde{\tau}_R$ case, mesons produced in $\tau$
decay may, in principle, induce showers before they decay, as
discussed above.  This contribution to hadronic energy is also shown
in Fig.~\ref{fig:branch}.  As can be seen, it is important only at
early times $\tau \alt 10^2 - 10^4~\s$ and may be safely ignored for
the decay times of most interest here.

Note that the hadronic branching fraction of \eqref{hadfraction} is
defined at the quark level.  Hadronization effects are important ---
for example, as we have shown above, mesons produced in $Z$ and $W$
decays do not contribute to hadronic cascades since they decay before
they interact.  Such effects have been included in deriving the
hadronic BBN constraints that we use~\cite{Kawasaki:2004yh}.

\begin{figure}
\resizebox{6.5 in}{!}{
\includegraphics{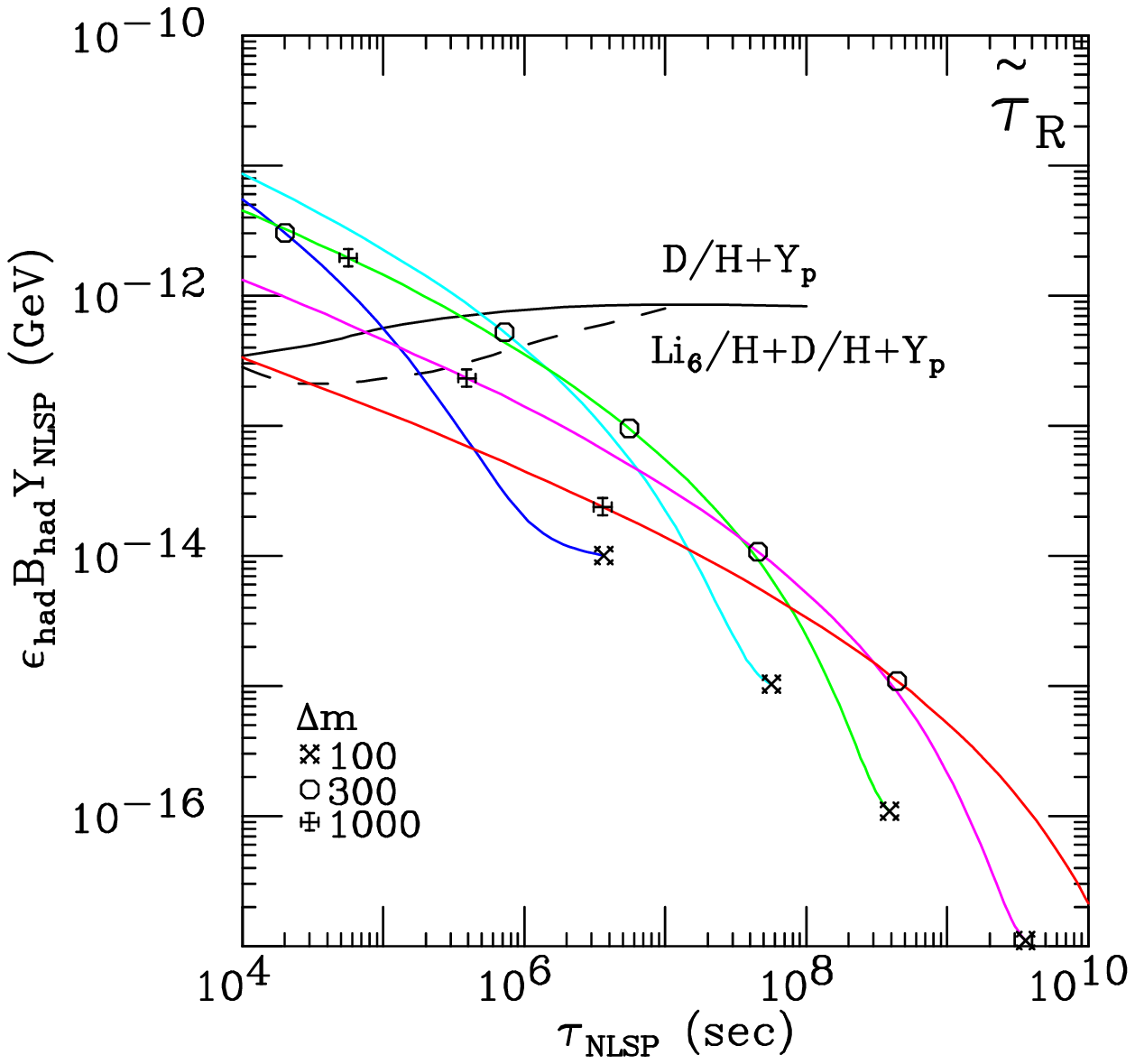}
\includegraphics{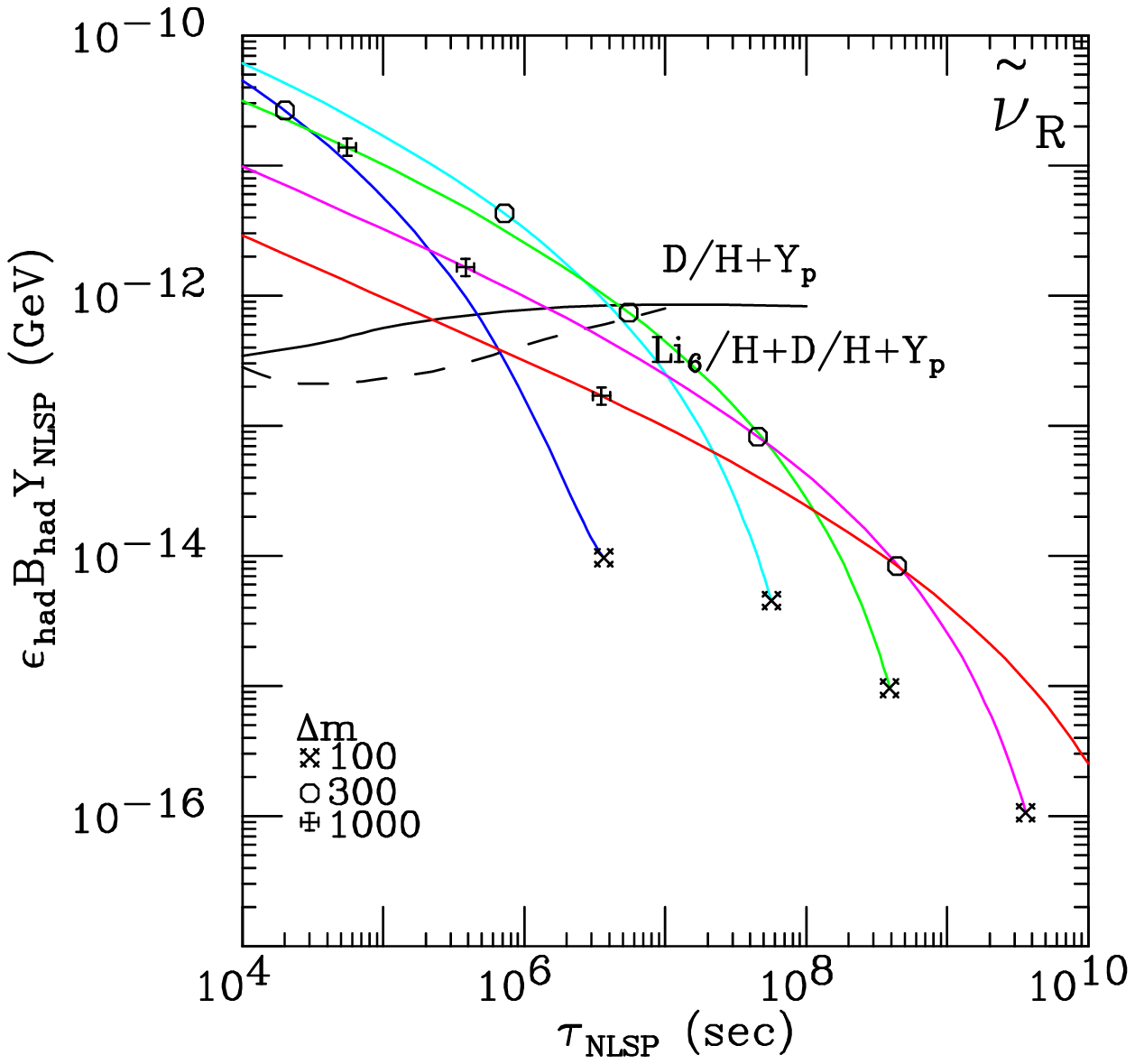}}
\caption{Hadronic energy releases $\xi_{\text{had}} \equiv
\epsilon_{\text{had}} B_{\text{had}} Y_{\text{NLSP}}$ as a function of
decay lifetime $\tau_{\text{NLSP}}$ for (left) $\tilde{\tau}_R$ NLSPs
and (right) $\tilde{\nu}$ NLSPs.  Along each curve, $\mgravitino$ is
held fixed and $\Delta m \equiv m_{\text{NLSP}} - \mgravitino$ varies.
Particular values of $\Delta m$ are marked by the symbols indicated.
The gravitino mass is $m_{\tilde{G}}=10, 10^2, 10^3, 10^4$ and $10^5$
GeV for the curves ordered from left to right by their value at
$\Delta m = 100~\gev$.  Constraints from BBN~\cite{Kawasaki:2004yh},
with and without ${}^6$Li/H, are also shown; regions above the curves
are disfavored.  We have assumed $\OmegaSWIMP = 0.23$ and
$\epsilon_{\text{had}}=\frac{1}{3} (m_{\text{NLSP}}-m_{\tilde{G}})$.
}
\label{fig:grid} 
\end{figure}

Fig.~\ref{fig:grid} shows the hadronic energy injection
$\xi_{\text{had}} \equiv \epsilon_{\text{had}} B_{\text{had}}
Y_{\text{NLSP}}$ as a function of decay lifetime $\tau_{\text{NLSP}}$
for $\tilde{\tau}_R$ and $\tilde{\nu}$ NLSPs.  Along each curve,
$\mgravitino$ is held fixed and $\Delta m \equiv m_{\text{NLSP}} -
\mgravitino$ varies.  The curves are truncated at $\Delta m =
100~\gev$ because for smaller $\Delta m$, three-body hadronic decays
are highly suppressed, leading to negligible hadronic energy
injection.  Hadronic BBN constraints from Ref.~\cite{Kawasaki:2004yh}
are also shown, with and without $^6$Li; the regions above the curves
are disfavored.  It is clear that for larger $m_{\tilde{G}}$, larger
values of $\Delta m$ becomes viable.  Constraints are stronger in the
$\tilde{\nu}$ case than in the $\tilde{\tau}_R$ case because the
hadronic branching fraction is larger for left-handed particles than
for right-handed ones.  Notice that we have fixed $\OmegaSWIMP=0.23$
in these plots.  The lines could be simply rescaled for other values
of $\OmegaSWIMP$ since $Y_{\text{NLSP}} \propto \OmegaSWIMP$.  We have
also adopted the hadronic BBN analysis of Ref.~\cite{Kawasaki:2004yh}.
If updated analyses of the BBN constraints become available, it is
straightforward to impose it in these plots to find out the remaining
viable parameter regions for $m_{\tilde{G}}$ and $\Delta m$.

\begin{figure}
\resizebox{6.5 in}{!}{
\includegraphics{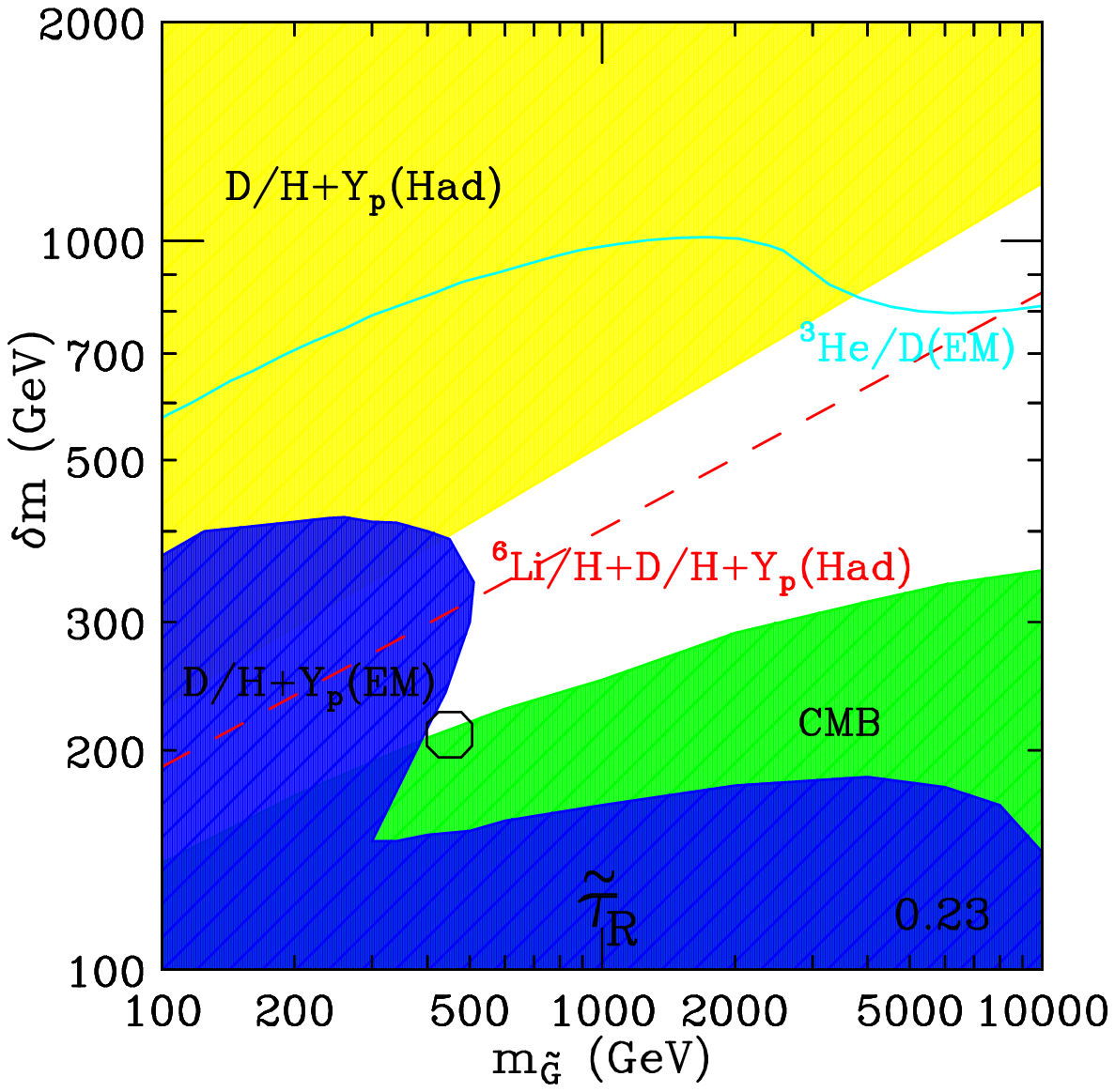}
\includegraphics{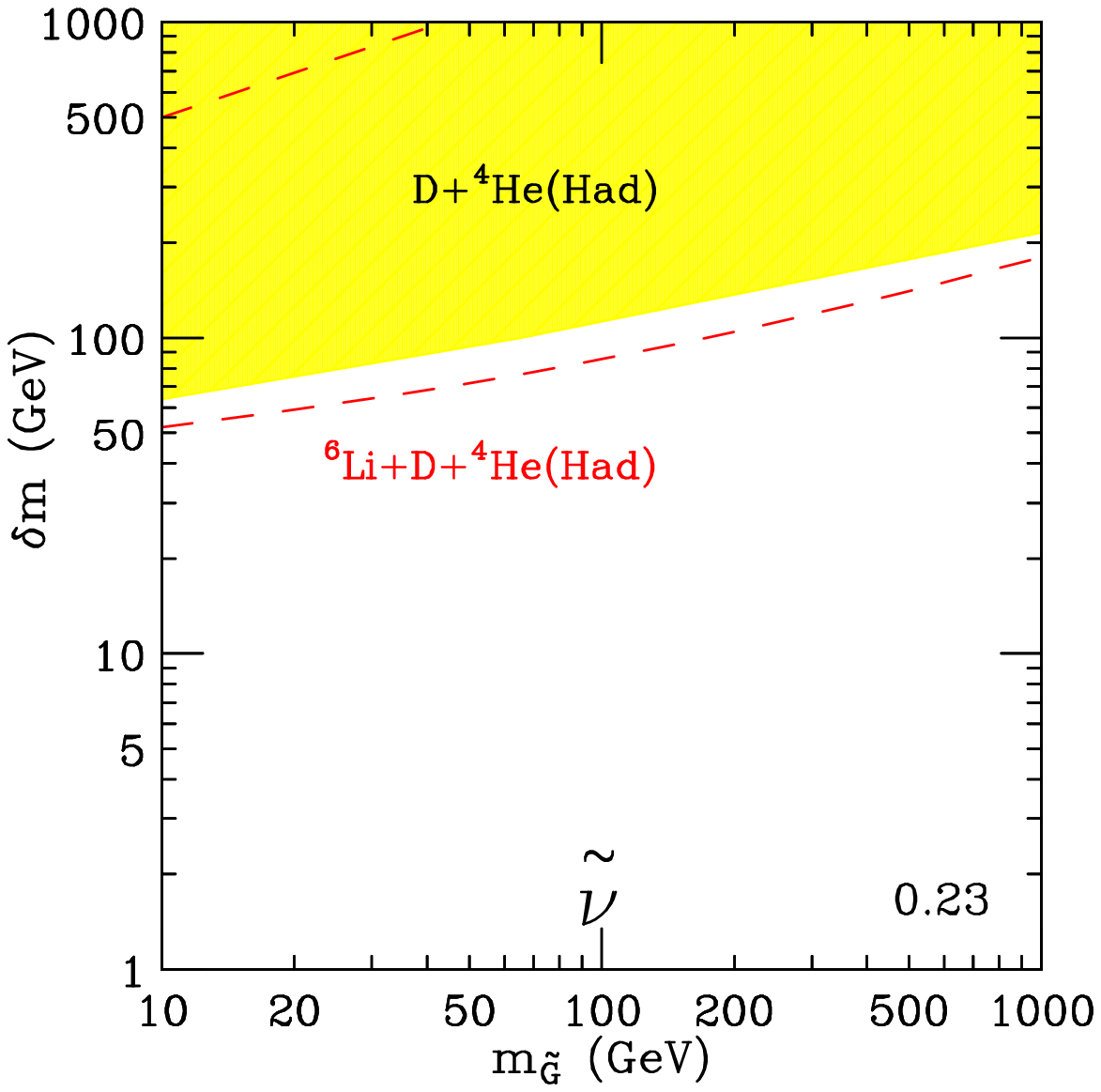}
}
\caption{Allowed and disfavored regions of the $(\mgravitino, \delta
m)$ plane, where $\delta m \equiv m_{\text{NLSP}} - \mgravitino -
m_Z$, for (left) $\tilde{\tau}_R$ and (right) $\tilde{\nu}$ NLSPs.
The shaded regions are disfavored by CMB, EM BBN, and hadronic BBN
constraints, as indicated.  The lines correspond to EM constraints
from ${}^3$He/D (solid) and hadronic constraints from ${}^6$Li/H
(dashed).  The circle indicates the best fit region where the $^7$Li
discrepancy is resolved.  See text for details.  We have assumed
$\OmegaSWIMP = 0.23$ and $\epsilon_{\text{had}}=\frac{1}{3}
(m_{\text{NLSP}}-m_{\tilde{G}})$.  Note that the axes have different
scales in the left and right panels.
\label{fig:summary} }
\end{figure}

Fig.~\ref{fig:summary} shows the allowed and disfavored regions of the
$(m_{\tilde{G}}, \delta m)$ plane for the two NLSP cases.  The CMB
$\mu$ distortion constraint, the EM BBN constraint of
Ref.~\cite{Cyburt:2002uv}, and the hadronic BBN constraint of
Ref.~\cite{Kawasaki:2004yh} are included.  For the $\tilde{\tau}_R$
case, the CMB constraint excludes a region with $\delta m \alt 150-
300~\gev$.  EM BBN constraints from D/H$+Y_p$ disfavor regions of
$\delta m$ below 400 GeV for $m_{\tilde{G}}\alt 500~\gev$ and $\delta
m$ below about 200 GeV for larger $m_{\tilde{G}}$.  Hadronic BBN
constraints from D/H$+Y_p$ disfavor the part of the parameter space
with large $\delta m$ where the slepton decays relatively early.  Note
that the hadronic constraint is extremely important --- it disfavors a
large and natural part of parameter space that would otherwise be
allowed.

Even after all of these constraints, however, the unshaded area in the
region $m_{\tilde{G}} \agt 200~\gev$ and $200~\gev \alt \delta m \alt
1500~\gev$ still remains viable.  The $\tilde{\tau}_R$ mass must be
above 500 GeV, but this is within reach of the LHC.  The ``best fit''
point where the $^7$Li discrepancy is fixed by EM energy release from
late slepton decays is indicated by a circle at $(m_{\tilde{G}},
\delta m) \approx (450~\gev, 200~\gev)$.  Note that it also remains
viable, even given the strong hadronic BBN constraints.

We also indicate the region that is disfavored by the less solid, but
stronger, constraints from ${}^3$He/D and ${}^6$Li/H.  The ${}^3$He/D
constraint excludes parameter space below the solid line.  The allowed
region would then be pushed up to larger $m_{\tilde{G}} \agt
3500~\gev$ and larger $\delta m \agt 800~\gev$. Such a heavy stau NLSP
is beyond the reach of the next generation hadron colliders.  On the
other hand, taking the ${}^6$Li/H constraint literally would exclude
parameter space above the dashed line.  Taking both constraints
literally and combining them would exclude most of the region of
parameter space shown.  Clearly, a more firm understanding of BBN will
have important consequences for the stau NLSP superWIMP scenario.

We have assumed $\OmegaSWIMP = 0.23$; of course, the allowed parameter
space would be enlarged if $\OmegaSWIMP < 0.23$.  We have also taken
$\epsilon_{\text{had}}=\frac{1}{3} (m_{\text{NLSP}} - m_{\tilde{G}})$
in our analysis; the allowed region could shift if there are
significant corrections to this relation.

The sneutrino NLSP case is also shown in Fig.~\ref{fig:summary}.
Hadronic BBN constraints from the D/H and $Y_p$ abundances only
disfavor $\delta m \agt$ 100 GeV. Even including the stronger but more
speculative ${}^6$Li/H constraint, there is still a large region of
$(\mgravitino, \delta m)$ that is allowed.  The origin of this result
is that at $\tau \agt 10^8~\s$, the sneutrino NLSP is completely free
of hadronic BBN constraints.  We have again assumed $\OmegaSWIMP =
0.23$.  To get this relic density from a sneutrino thermal relic
density, one would typically require sneutrino masses above 500 GeV.
However, even assuming a thermal relic abundance, if the superWIMP
component is only part of the present dark matter, smaller sneutrino
masses are allowed.

For right-handed selectron and smuon NLSPs, the results will be
similar to the right-handed stau NLSP case.  The EM constraints are
slightly stronger, since all the released energy induces EM cascades.
For left-handed slepton NLSPs, the region of $(m_{\tilde{G}}, \delta
m)$ that is disfavored by hadronic injection is similar to the region
disfavored for the sneutrino NLSP.  In addition, however, there are EM
BBN constraints that are similar to those for right-handed slepton
NLSPs.

\section{Conclusions}
\label{sec:conclusion}

In this paper, we explored the possibility of superWIMPs as candidates
for dark matter.  Examples of superWIMPs are the gravitino in
supersymmetric models and the lightest Kaluza-Klein 
graviton in universal extra
dimension models.  SuperWIMPs obtain their relic density through late
decays: WIMP $\to$ superWIMP $+$ SM particles.  The decay usually
occurs between $10^4$ s and $10^8$ s, which has important cosmological
implications.  Such late decays release EM and hadronic energy into
the Universe, which may affect BBN predictions for the primordial
abundance of the light elements.  The constraints on EM injection have
been studied in detail in Ref.~\cite{Feng:2003uy}.  In this paper, we
have analyzed the hadronic BBN constraints.  We have taken the
lightest gravitino as a concrete example of a superWIMP and have
focused on slepton and sneutrino NLSPs as the most promising WIMP
candidates.

We have determined the hadronic energy release by calculating
three-body decay widths in detail.  For the cases of $\tilde{\tau}_R$
and $\tilde{\nu}$ NLSPs, the hadronic decay branching fraction is
below the level of $10^{-3}$ when $m_{\tilde{\tau}_R} \alt 1~\tev$ and
$m_{\tilde{\nu}} \alt 300~\gev$, respectively, or when the decay
lifetime is $\tau \agt 10^6~\s$.  We identified the allowed and
disfavored regions of the $(m_{\tilde{G}}, \delta m)$ plane, imposing
CMB constraints, EM BBN constraints, and hadronic BBN constraints.
For the sneutrino NLSP case, $\delta m \alt 60 - 200~\gev$ is allowed
for a large range of $m_{\tilde{G}}$.  For the stau NLSP, the hadronic
constraints are weaker: $\delta m \alt 200-1200~\gev$.  However,
additional constraints from CMB and EM energy injection apply.
Combining all the constraints for the right-handed stau NLSP, the
allowed window is $300~\gev \alt m_{\tilde{G}} \alt 1~\tev$, $200~\gev
\alt \delta m \alt 1200~\gev$, corresponding to $m_{\tilde{\tau}_R}
\agt 500~\gev$.  We have assumed $\OmegaSWIMP = 0.23$ in our
analysis.  The constraints would be relaxed if superWIMPs are only
part of the present dark matter.  In addition, there are still
ambiguities in the BBN constraints if the decay time lies in the
region where EM and hadronic effects are comparable.  In particular,
as we discussed, their effects on D might cancel, and the allowed
parameter space could be enlarged.  Progress in firming up BBN
constraints in this region, and for other elements, such as $^3$He and
$^6$Li may have crucial implications for the superWIMP dark matter
scenario.

Although the superWIMP itself would escape all direct and indirect
dark matter searches, the slepton/sneutrino NLSP will have rich
implications for collider phenomenology.  Such metastable sleptons
will not decay inside the detector, resulting in signals of highly
ionizing tracks for sleptons and missing energy signals for
sneutrinos.  Discussion of the collider phenomenology, combined with
thermal relic density calculations for the NLSP in supergravity models
will be presented in Ref.~\cite{Feng:2004}.

Although we have focused on the SUSY scenario in this paper,
superWIMPs could in general be any gravitationally interacting
particle that obtains its relic density through the late decay of a
WIMP.  The discussion and results for late decay of a leptonic WIMP in
other models will be qualitatively similar to the discussion and
results presented here.  This scenario also suggests that a particle
that appears to play the role of dark matter at late times, even after
BBN, could very well be different from the particle that constitutes
dark matter now.  The very late decay of ``would be'' dark matter
particles may have important cosmological implications, for example,
affecting small scale structure.  This feature may provide
qualitatively new possibilities for explaining puzzling cosmological
observations~\cite{Chen:2003gz,Sigurdson:2003vy}.
  
We have assumed $R$-parity conservation in our discussion.  In the
case of $R$-parity violation (RPV), the gravitino could still
constitute dark matter as long as its decay lifetime is comparable to
or longer than the age of
Universe~\cite{Takayama:2000uz,Moreau:2001sr}.  Gravitino dark matter
in an RPV scenario could be distinguished from the superWIMP scenario
by both cosmological observations and collider experiments.  For
example, the decay of even a tiny amount of gravitino dark matter into
SM particles in an RPV scenario could be seen in the diffuse photon
flux.  In addition, if the RPV is not extremely small, the collider
signatures could be different from those in the superWIMP scenario.
Further work is needed to study how to distinguish these two
scenarios.

\begin{acknowledgments}
We are grateful to M.~Fujii, K.~Kohri, T.~Moroi, and A.~Rajaraman for
helpful conversations. The work of JLF was supported in part by
National Science Foundation CAREER Grant PHY--0239817, and in part by
the Alfred P.~Sloan Foundation.
\end{acknowledgments}

\appendix*

\section{Widths for Three-Body Slepton Decays}

\subsection{Interactions}

The gravitino-gaugino-gauge boson interaction is
\begin{eqnarray}
{\cal L} = - \frac{i}{8M_*}
\overline{\psi}_{\mu} \left[ \gamma^{\nu}, \gamma^{\rho} \right]
\gamma^{\mu}\lambda F_{\nu\rho} \ ,
\end{eqnarray}
where $M_* = \sqrt{2} M_4 = 2.4 \times 10^{18}~\gev$ is the reduced
Planck mass, and $\psi_{\mu} \equiv \psi_{\mu}(x)$ denotes the
gravitino field at spacetime point $x$.  The gravitino-slepton-lepton
interaction is
\begin{eqnarray}
{\cal L} = - \frac{1}{\sqrt{2} M_*} \partial_{\nu} \tilde{l} \, 
\overline{l_h} \, \gamma^{\mu} \gamma^{\nu} \psi_{\mu} \ ,
\end{eqnarray}
where subscript $h=L,R$ denotes the lepton chirality.  In addition,
there are gravitino-slepton-lepton-gauge boson interactions:
\begin{equation}
{\cal L} = \frac{ie g_V^{l_h}}{\sqrt{2} M_*} A_{\nu} \tilde{l} \, 
\overline{l_h} \, \gamma^{\mu} \gamma^{\nu} \psi_{\mu} \ .
\end{equation}
Here $g_V^{l_h}$ is the gauge coupling coefficient, which is 
given explicitly below.

Since gravitinos are Majorana particles,
\begin{eqnarray}
\psi_{\mu}^c \equiv C \overline{\psi}_{\mu}^T
= \psi_{\mu} \ .
\end{eqnarray}
The gravitino spin sum (with four momentum $p_{\mu}$) is
\begin{eqnarray}
\sum_h {\tilde{G}^h_{\mu}}(p) \overline{\tilde{G}^h_{\nu}}(p)
&=& - \left(\gamma \cdot p + m_{\tilde{G}}\right) \left(g_{\mu\nu}
- \frac{p_{\mu}p_{\nu}}{m_{\tilde{G}}^2}\right) \nonumber \\
&& - \frac{1}{3}\left(\gamma_{\mu} + \frac{p_{\mu}}{m_{\tilde{G}}}\right)
\left(\gamma \cdot p - m_{\tilde{G}}\right)
\left(\gamma_{\nu} + \frac{p_{\nu}}{m_{\tilde{G}}}\right) \ .
\end{eqnarray}

\subsection{Matrix Amplitudes}

Here we present the matrix amplitudes for the three-body decays
$\tilde{l} \to l' V \tilde{G}$, where $l'$ is the SM partner of
$\tilde{l}$ or its weak isospin partner, and $V=Z$ or $W$.  The
Feynman diagrams are shown in Fig.~\ref{fig:feynmangraphs}.

We define the following ten operators:
\begin{eqnarray}
{\cal O}_1 &=& \overline{l'_h}\ p_{\tilde{l}}\cdot\tilde{G}^c\ 
p_{l'}\cdot\epsilon^* \\
{\cal O}_2 &=& \overline{l'_h}\ p_l\cdot\tilde{G}^c \ 
p_{\tilde{l}}\cdot\epsilon^* \\
{\cal O}_3 &=& \overline{l'_h}\ p_{\tilde{G}}\cdot\tilde{G}^c
\ p_{\tilde{l}}\cdot\epsilon^* \\
{\cal O}_4 &=&i\ \overline{l'_h}\ \gamma\cdot\epsilon^*\ 
p_V\cdot \tilde{G}^c \\
O_5 &=&\overline{l'_h}\ \gamma\cdot p_V\ 
\gamma\cdot\epsilon^*\ p_{\tilde{l}} \cdot \tilde{G}^c \\
{\cal O}_6 &=&i\ \overline{l'_h}\ \gamma\cdot p_V
\ \gamma\cdot\epsilon^* \ p_V \cdot \tilde{G}^c \\
{\cal O}_7 &=& i\ \overline{l'_h}\ \gamma\cdot p_{\tilde{G}}\ 
\gamma\cdot\epsilon^* \ p_V \cdot \tilde{G}^c \\
{\cal O}_8 &=& i \ \overline{l'_h}\ \epsilon^*\cdot\tilde{G}^c \\
{\cal O}_9 &=& i \ \overline{l'_h}\  \gamma\cdot p_V\ 
\epsilon^*\cdot \tilde{G}^c \\
{\cal O}_{10} &=& i \ \overline{l'_h}\ \gamma\cdot p_{\tilde{G}}
\ \gamma\cdot p_V \epsilon^*\cdot\tilde{G}^c\ ,
\end{eqnarray}
where 
$\epsilon^*_\mu$ is the polarization of gauge boson $V$.  Notice that
for an on-shell gravitino in the final state, ${\cal O}_3=0$ by using
the gravitino equation of motion.

In terms of these operators, the matrix elements for three-body
slepton decay [diagrams (a)--(d) in Fig.~\ref{fig:feynmangraphs}] are
\begin{eqnarray}
{\cal {M}}_V^a &=& \frac{eg_V^{l_h}}{\sqrt{2}M_*}
\frac{1}{m_{23}^2} \left(2{\cal O}_1-{\cal O}_5 \right)
\label{eq:meleZ1}\\
{\cal M}_V^b &=& \frac{eg_V^{l_h}}{\sqrt{2}M_*}
\frac{1}{m^2_{\tilde{G}}+m_V^2-m^2_{13}-m^2_{23}} 
\left({\cal O}_2+{\cal O}_3 \right)
\label{eq:meleZ2}\\
{\cal {M}}_V^c &=& \sum_i \frac{eg_{\chi_i}^{l_h} 
g_{\chi_i V}^h}{4M_*} \frac{1}{m^2_{13}-m^2_{\chi_i}}
\left[m_{\chi_i} \left({\cal O}_4-{\cal O}_9 \right)
- 4 \left({\cal O}_6+{\cal O}_7 - {\cal O}_{10} \right) 
+ m_V^2{\cal O}_8 \right]
\label{eq:meleZ3}\\
{\cal {M}}_V^d &=& i \frac{\sqrt{2}eg_V^{l_h}}{M_*}{\cal O}_8 \ , 
\label{eq:meleZ4}
\end{eqnarray}
where $\chi_i$ denotes a neutralino or chargino, the couplings
$g_V^{l_h}$, $g_{\chi_i}^{l_h}$, and $g_{\chi_i V}$ are given below
for each decay process, and
\begin{eqnarray}
m^2_{12}=(p_{\tilde{G}}+p_{l'})^2 \ , \quad
m^2_{13}=(p_{\tilde{G}}+p_V)^2 \ , \quad 
m^2_{23}=(p_{l'}+p_V)^2 \ .
\end{eqnarray}
Notice that $m_{12}^2+m_{13}^2+m_{23}^2=\msl^2+\mg^2+m_Z^2$.  Given
this relation, the four-momentum scalar products that appear below in
the expressions for squared matrix elements can be expressed in terms
of $m_{12}^2$, $m_{13}^2$ and $m_{23}^2$.

For each specific decay, the couplings in
Eqs.~(\ref{eq:meleZ1})--(\ref{eq:meleZ4}) should be replaced according
to the following rules:

\begin{itemize}
\item{$\tilde{l}_L\to lZ\tilde{G}$}
\begin{eqnarray}
h &\to& L\ , \qquad V\to Z\\ 
g_V^{l_h} &\to& g_Z^{l_L}
= \frac{1}{\sin \theta_W\cos \theta_W} 
\left(\frac{1}{2}-\sin^2 \theta_W \right)\\
g_{\chi_i}^{l_h} &\to& g_{\chi^0_i}^{l_L}
=\frac{1}{\sqrt{2}\sin \theta_W\cos \theta_W}
(N_{i2}^*\cos \theta_W+N_{i1}^*\sin \theta_W)\\
g_{\chi_i V}^h &\to& g_{\chi_i Z}^L
= N_{i1}^*(-\sin \theta_W)+N_{i2}^*(\cos \theta_W)
\end{eqnarray}
\item{$\tilde{l}_L\to \nu W \tilde{G}$ 
and $\tilde{\nu}_L \to l W \tilde{G}$}
\begin{eqnarray}
h &\to& L\ , \qquad V\to W \\ 
g_V^{l_h} &\to& g_W^{l_L}=-\frac{1}{\sqrt{2}\sin \theta_W}\\
g_{\chi_i}^{l_h} &\to& g_{\chi^-_i}^{l_L}
=\frac{1}{\sqrt{2}\sin \theta_W}V_{i1}^*\\
g_{\chi_i V}^h &\to& g_{\chi_i W}^L = V_{i1}^*
\end{eqnarray}
\item{$\tilde{l}_R\to l Z \tilde{G}$}
\begin{eqnarray}
h &\to& R\ , \qquad V\to Z\\
g_V^{l_h} &\to& g_Z^{l_L}
=\frac{1}{\sin \theta_W\cos \theta_W} \left( - \sin^2 \theta_W \right) \\
g_{\chi_i}^{l_h} &\to& g_{\chi^0_i}^{l_R}
= -\frac{\sqrt{2}}{\cos \theta_W} N_{i1}\\
g_{\chi_i V}^h &\to& g_{\chi_i Z}^R
=N_{i1}(-\sin \theta_W)+N_{i2}(\cos \theta_W)
\end{eqnarray}
\item{$\tilde{\nu}_L \to \nu Z \tilde{G}$}
\begin{eqnarray}
h &\to& L\ , \qquad V\to Z\\
g_V^{l_h} &\to& g_Z^{l_L} = \frac{1}{\sin \theta_W\cos \theta_W} 
\left( -\frac{1}{2} \right) \\
g_{\chi_i}^{l_h} &\to& g_{\chi^0_i}^{l_L}
= \frac{1}{\sqrt{2}\sin \theta_W\cos \theta_W}
(N_{i2}^*\cos \theta_W+N_{i1}^*\sin \theta_W)\\
g_{\chi_i V}^h &\to& g_{\chi_i Z}^L
= N_{i1}^*(-\sin \theta_W)+N_{i2}^*(\cos \theta_W)
\end{eqnarray}
\end{itemize}
In these expressions, $V$ and $N$ are matrices that diagonalize the
chargino and neutralino mass matrices, following the conventions of
Ref.~\cite{Haber:1984rc}.

\subsection{Squared Matrix Elements}

The differential decay width is
\begin{eqnarray}
 d\Gamma=\frac{1}{(2\pi)^3}\frac{1}{32m^3_{\tilde{l}}}
 |{\cal M}|^2 dm^2_{13}dm^2_{23} \ ,
\end{eqnarray}
where ${\cal M}={\cal M}_V^a+{\cal M}_V^b+{\cal M}_V^c+{\cal M}_V^d$.  

The sum of the matrix elements can be written as
\begin{eqnarray}
{\cal{M}}(\tilde{l}_h\to l' V \tilde{G})
= {\cal M}_V^a+{\cal M}_V^b+{\cal M}_V^c + {\cal M}_V^d
= \sum_{i=1\ldots 10} {\cal{M}}_i \calO_i \ ,
\end{eqnarray}
where the ${\cal{M}}_i$ can be read off from
Eqs.~(\ref{eq:meleZ1})--(\ref{eq:meleZ4}) above.  The squared matrix
element is
\begin{eqnarray}
\left| {\cal{M}}(\tilde{l}_h \to l' V \tilde{G}) \right|^2 
&=& \sum_{i=1\ldots 10} |{\cal{M}}_i|^2 \calO_{i,i} \nonumber \\
&& + \sum_{i,j=1\ldots 10}^{i<j} \Real({\cal{M}}_i {\cal{M}}_j^*) 
\calO_{i,j}^{\text re} 
+ \sum_{i,j=1\ldots 10}^{i<j} \Imag({\cal{M}}_i {\cal{M}}_j^*) 
\calO_{i,j}^{\text im} \ .
\label{eq:square}
\end{eqnarray}
In our calculations, we chose the convention that the diagonalizing
matrices $V$ and $N$ are real. All couplings appearing in
Eq.~(\ref{eq:meleZ1})$-$(\ref{eq:meleZ4}) are therefore real.  Thus,
all the ${\cal{M}}_i$ are real except for ${\cal M}_8$, which has both
real and imaginary components.  The only non-zero contributions to the
last term in Eq.~(\ref{eq:square}) come from $\Imag({\cal{M}}_i
{\cal{M}}_8^*)$ ($i<8$) and $\Imag({\cal{M}}_8 {\cal{M}}_j^*)$
($j>8$).  Expressions for $\calO_{i,i}$, $\calO_{i,j}^{\text re}$,
$\calO_{i,8}^{\text im}$ ($i<8$) and $\calO_{8,j}^{\text im}$ ($j>8$)
are:
\begin{eqnarray}
\calO_{1,1} &=& -\frac{4}{3\mg^2m_V^2}
\left[\mg^2\msl^2 - \left(\pgpsl\right)^2\right] 
\left[\mg^2 - \pgpsl + \pgpZ\right] \nonumber\\
&& \!\!\!\!\!\!\!\! \times \left[\mg^2m_V^2 + \msl^2m_V^2 - 2m_V^2\pgpsl 
- \left(\pgpZ\right)^2 + 2\pgpZ\pslpZ
- \left(\pslpZ\right)^2\right] 
\\
\calO_{2,2}&=& -\frac{4}{3\mg^2m_V^2}
\left[\mg^2 - \pgpsl + \pgpZ\right]
\left[\msl^2m_V^2 - \left(\pslpZ\right)^2\right] \nonumber \\ 
&& \!\!\!\!\!\!\!\! \times \left[-\left(\pgpsl\right)^2 
+ 2\pgpsl\pgpZ - \left(\pgpZ\right)^2 +  \mg^2\left(\msl^2 + m_V^2
- 2\pslpZ \right) \right] 
\\
\calO_{3,3} &=& {{0}} 
\\
\calO_{4,4} &=&
\frac{4}{3\mg^2m_V^2}
\left[\mg^2m_V^2 - \left(\pgpZ\right)^2\right] \nonumber \\
&& \times \left[\mg^2m_V^2 - m_V^2\pgpsl + 2\left(\pgpZ\right)^2 
+ \pgpZ \left(3m_V^2 - 2\pslpZ\right)\right]
\\
\calO_{5,5} &=& -\frac{4}{3\mg^2}
\left[\mg^2\msl^2 - \left(\pgpsl\right)^2\right] \nonumber \\
&& \times
\left[\mg^2m_V^2 - m_V^2\pgpsl - 4\left(\pgpZ\right)^2 + \pgpZ
      \left(-3m_Z^2 + 4\pslpZ\right)\right]
\\
\calO_{6,6} &=& -\frac{4}{3\mg^2}
\left[\mg^2m_Z^2 - \left(\pgpZ\right)^2\right] \nonumber \\
&& \times
\left[\mg^2m_V^2 - m_V^2\pgpsl -  4\left(\pgpZ\right)^2 + \pgpZ
\left(-3m_V^2 + 4\pslpZ\right)\right]
\\
\calO_{7,7} &=& \frac{4}{3\mg^2m_V^2}
\left[ \mg^2m_V^2 - \left(\pgpZ\right)^2\right] \nonumber \\
&& \left\{ \mg^4m_V^2 + 2\mg^2\left(\pgpZ\right)^2 
+ 4\pgpZ^3 - \pgpsl \left[ \mg^2m_V^2 
+ 4\left(\pgpZ\right)^2\right] \right. \nonumber\\
&& \left. - \mg^2\pgpZ \left(m_V^2 - 2\pslpZ\right)\right\}
\\
\calO_{8,8} &=&-\frac{4}{3\mg^2m_V^2}
\left[\mg^2 - \pgpsl + \pgpZ\right]
\left[2\mg^2m_V^2 + \left(\pgpZ\right)^2\right]
\\
\calO_{9,9} &=& \frac{4}{3\mg^2m_V^2}
\left[ 2\mg^2m_V^2 + \left(\pgpZ\right)^2\right] \nonumber \\
&& \times
\left[\mg^2m_V^2 - m_V^2\pgpsl -  2\left(\pgpZ\right)^2 - \pgpZ
      \left(m_V^2 - 2\pslpZ\right)\right]
\\
\calO_{10,10}&=&\frac{4}{3\mg^2m_V^2} 
\left[2\mg^2m_V^2 + \left(\pgpZ\right)^2\right] \nonumber \\ 
&& \left\{ \mg^4m_V^2 - 2\mg^2\left(\pgpZ\right)^2 
- 4\left(\pgpZ\right)^3 \right. \nonumber\\
&& \left. + \pgpsl\left[-\mg^2m_V^2 
+ 4\left(\pgpZ\right)^2\right] 
+ \mg^2\pgpZ\left(3m_V^2 -  2\pslpZ\right)\right\}
\\
\calO_{1,2}^{\text re}&=&\frac{8}{3\mg^2m_V^2} 
\left[\mg^2 - \pgpsl + \pgpZ \right] \nonumber \\
&& \times \left[ - \left(\pgpsl\right)^2 + \pgpsl \pgpZ 
+ \mg^2\left(\msl^2 - \pslpZ\right)\right] \nonumber \\ 
&&\times \left[ -\msl^2m_V^2 + m_Z^2\pgpsl -
\pgpZ \pslpZ + \left(\pslpZ\right)^2\right]
\\
\calO_{1,3}^{\text re} &=& \calO_{1,4}^{\text re} = {{0}} 
\\
\calO_{1,5}^{\text re} &=& \frac{8}{3\mg^2}
\left[\mg^2\msl^2 - \left(\pgpsl\right)^2\right]
\left[ -\left(\pgpZ\right)^2 + 
\mg^2\left(m_V^2 - \pslpZ\right) \right. \nonumber\\
&&
\left. +  \pgpZ\left(-\msl^2 + \pslpZ\right) 
+ \pgpsl \left(-m_V^2 + \pgpZ + \pslpZ\right)\right]
\\
{\cal{O}}_{1,6}^{\text re} &=& {\cal{O}}_{1,7}^{\text
re}={\cal{O}}_{1,8}^{\text re}={\cal{O}}_{1,9}^{\text
re}={\cal{O}}_{1,10}^{\text re}=0
 \\
{\cal{O}}_{2,3}^{\text re}&=&{\cal{O}}_{2,4}^{\text re}={{0}}
 \\
{\cal{O}}_{2,5}^{\text re}&=&\frac{4}{3\mg^2} 
\left\{
\left(\pgpsl\right)^2\left[\mg^2m_V^2 
+ 2\left(\msl^2- m_V^2\right)\pgpZ\right] \right.
\nonumber \\ &&
+ 2\left(\pgpsl\right)^3\left(m_V^2 - \pslpZ\right) 
- 2\pgpsl \left[ \left(\pgpZ\right)^2 
\left(\msl^2 - \pslpZ\right) \right. \nonumber \\
&&\left. + \mg^2\left(\msl^2 -\pslpZ\right) 
\left(m_V^2 - \pslpZ\right) +  \mg^2\pgpZ\pslpZ\right] 
\nonumber \\ &&
+\mg^2\left[\msl^2\left(\pgpZ\right)^2 
- 2\pgpZ \left(\msl^2 - \pslpZ\right)^2 \right. \nonumber \\
&& \left. \left. + \mg^2\left(-\msl^2m_V^2 
+ \left(\pslpZ\right)^2\right)\right]\right\}
 \\
{\cal{O}}_{2,6}^{\text re}&=&{\cal{O}}_{2,7}^{\text
re}={\cal{O}}_{2,8}^{\text re}={\cal{O}}_{2,9}^{\text
re}={\cal{O}}_{2,10}^{\text re}={{0}}
 \\
{\cal{O}}_{3,4}^{\text re}&=&{\cal{O}}_{3,5}^{\text
re}={\cal{O}}_{3,6}^{\text re}={\cal{O}}_{3,7}^{\text
re}={\cal{O}}_{3,8}^{\text re}={\cal{O}}_{3,9}^{\text
re}={\cal{O}}_{3,10}^{\text re}={{0}}
 \\
{\cal{O}}_{4,5}^{\text re}&=&{{0}}
 \\
{\cal{O}}_{4,6}^{\text re}&=& \frac{8}{\mg}
\left[\mg^2m_V^2 - \left(\pgpZ\right)^2\right] \left[m_V^2 +
     \pgpZ - \pslpZ\right]
 \\
\calO_{4,7}^{\text re}&=& \frac{8}{\mg}
\left[\mg^2 - \pgpsl + \pgpZ\right]
\left[\mg^2m_V^2 - \left(\pgpZ\right)^2\right]
 \\
\calO_{4,8}^{\text re}&=& \frac{8}{3\mg m_V^2} 
\left[-m_V^2\left(\mg^2 - 2\pgpsl\right)\pgpZ + \pgpZ^ 3 \right. \nonumber\\
&&
\left. + \mg^2m_V^2\left(m_V^2 - \pslpZ\right) -
       \left(\pgpZ\right)^2\left(m_V^2 + \pslpZ\right)\right]
 \\
\calO_{4,9}^{\text re}&=& \frac{8}{3\mg^2 m_V^2} 
 \left[\mg^4m_V^4 + \mg^2m_V^4\pgpZ + 
      \mg^2m_V^2\left(\pgpZ\right)^2 -  2\left(\pgpZ\right)^4 \right.
\nonumber \\ &&
\left. -  m_V^2\pgpsl\left(\mg^2m_V^2 +  \left(\pgpZ\right)^2\right) 
- \left(\pgpZ\right)^3 \left(m_V^2 - 2\pslpZ\right)\right]
\\
\calO_{4,10}^{\text re}&=&
\frac{8}{3\mg}
\left\{\mg^4m_V^2 - \mg^2\left(\pgpZ\right)^2 - 
      3\pgpZ^3 +  \pgpsl\left[-\mg^2m_V^2 +
      3\left(\pgpZ\right)^2\right] \right. \nonumber\\
&& \left. + \mg^2\pgpZ\left(3m_V^2 -  2\pslpZ\right)\right\}
\\
{\cal{O}}_{5,6}^{\text re}&=&{\cal{O}}_{5,7}^{\text
re}={\cal{O}}_{5,8}^{\text re}={\cal{O}}_{5,9}^{\text
re}={\cal{O}}_{5,10}^{\text re}={{0}}
 \\
\calO_{6,7}^{\text re}&=& \frac{8}{3\mg^2}
\left[\mg^2m_V^2 - \left(\pgpZ\right)^2\right] \nonumber \\
&& \times 
\left[\left(3\mg^2 - 2\pgpsl\right)\pgpZ
+ 2\left(\pgpZ\right)^2 + 
\mg^2\left(m_V^2 - \pslpZ\right)\right]
 \\
\calO_{6,8}^{\text re}&=&-\frac{8}{3\mg^2}
 \left[\mg^4m_V^2 - \mg^2\left(\pgpZ\right)^2 
+ \left(\pgpZ\right)^3 \right. \nonumber\\
&& \left. - \pgpsl \left(\mg^2m_V^2 + \left(\pgpZ\right)^2\right) 
- \mg^2\pgpZ \left(m_V^2 - 2\pslpZ\right)\right]
 \\ 
\calO_{6,9}^{\text re}&=&\frac{8}{3\mg}
\left[m_V^2\left(3\mg^2 - 2\pgpsl\right) \pgpZ - 3\left(\pgpZ\right)^3 
\right. \nonumber\\
&&
\left. - \left(\pgpZ\right)^2 \left(m_V^2 - 3\pslpZ\right) 
+ \mg^2m_V^2\left(m_V^2 - \pslpZ\right)\right]
\\
\calO_{6,10}^{\text re}&=&\frac{8}{3\mg^2}
\left[\mg^4m_V^2\pgpZ -  \left(\mg^2 - 2\pgpsl\right) \left(\pgpZ\right)^3 -
      2\left(\pgpZ\right)^4 
\right. \nonumber \\ 
&& \left. + \mg^4m_V^2\left(m_V^2 - \pslpZ\right)
+ \mg^2\left(\pgpZ\right)^2 \left(m_V^2 - \pslpZ\right)\right]
 \\
\calO_{7,8}^{\text re}&=&\frac{8}{3\mg^2m_V^2}
 \left[\mg^4m_V^2\pgpZ -  \left(\mg^2 - 2\pgpsl\right)\left(\pgpZ\right)^3 -
      2\left(\pgpZ\right)^4  
\right. \nonumber \\ &&
\left. + \mg^4m_V^2\left(m_V^2 - \pslpZ\right) 
+ \mg^2\left(\pgpZ\right)^2 \left(m_V^2 - \pslpZ\right)\right]
 \\
\calO_{7,9}^{\text re}&=&\frac{8}{3\mg}
\left[\mg^4m_V^2 - \mg^2\left(\pgpZ\right)^2 +  \left(\pgpZ\right)^3
\right. \nonumber\\
&&
\left. - \pgpsl \left(\mg^2m_V^2 + \left(\pgpZ\right)^2\right)
     -  \mg^2\pgpZ \left(m_V^2 - 2\pslpZ\right)\right]
 \\
\calO_{7,10}^{\text re}&=&
\frac{8}{3\mg^2m_Z^2}
\left\{\mg^6m_V^4 + \mg^4m_V^4\pgpZ +  \mg^4m_V^2\left(\pgpZ\right)^2 
\right. \nonumber \\ &&
-  2\mg^2\left(\pgpZ\right)^4      -  4\left(\pgpZ\right)^5
\nonumber \\
&& \left. -  \pgpsl\left[\mg^4m_V^4 +  \mg^2m_V^2\left(\pgpZ\right)^2 -
  4\left(\pgpZ\right)^4\right] \right.
\nonumber\\
&&
\left. + \mg^2\left(\pgpZ\right)^3 \left(3m_V^2 - 2\pslpZ\right)\right\}
 \\
\calO_{8,9}^{\text re}&=&\frac{8}{3\mg m_V^2}
\left[ 2\mg^2m_V^2 + \left(\pgpZ\right)^2\right] 
\left[m_V^2 + \pgpZ -\pslpZ\right]
 \\
\calO_{8,10}^{\text re}&=&\frac{8}{3\mg^2 m_V^2}
\left[2\mg^2m_V^2 + \left(\pgpZ\right)^2\right] \nonumber \\
&& \times
\left[-\left(\mg^2 - 2\pgpsl\right)\pgpZ 
- 2\left(\pgpZ\right)^2 +\mg^2\left(m_V^2 - \pslpZ\right)\right]
 \\
\calO_{9,10}^{\text re}&=&\frac{8}{3\mg}
\left[\mg^2 - \pgpsl+ \pgpZ\right]
\left[2\mg^2m_V^2 + \left(\pgpZ\right)^2\right] \\
\calO_{1,8}^{\text im}&=&
-\frac{8}{3\mg^2m_V^2}(\mg^2-\pgpsl+\pgpZ)
\left\{\left[\msl^2m_V^2-(\pslpZ)^2+(\pgpZ)(\pslpZ)\right]\mg^2
\right. \nonumber \\
&& \left. -m_V^2(\pgpsl)^2
+(\pgpsl)(\pgpZ)(\pslpZ-\pgpZ)\right\} \\
\calO_{2,8}^{\text im}&=&
-\frac{8}{3\mg^2m_V^2}(\mg^2-\pgpsl+\pgpZ)
\left[(\msl^2m_V^2-(\pslpZ)^2)\mg^2-m_V^2(\pgpsl)^2\right.
\nonumber \\
&& \left.-(\pgpZ)^2\pslpZ
+(\pgpsl)(\pgpZ)(m_V^2+\pslpZ)\right]
\\
\calO_{3,8}^{\text im}&=&\calO_{4,8}^{\text im}=0\\
\calO_{5,8}^{\text im}&=&
\frac{8}{3\mg^2}\left\{
(\pgpsl-\mg^2)\pslpZ \mg^2 -\pgpsl(\pgpZ)^2\right.
\nonumber \\
&& \left.
+\pgpZ\left[\pgpsl \mg^2
+(\pslpZ-2 \msl^2)\mg^2+(\pgpsl)^2\right]\right\}
\\
\calO_{6,8}^{\text im}&=&\calO_{7,8}^{\text im}
=\calO_{8,9}^{\text im}=\calO_{8,10}^{\text im}=0
\ .
\end{eqnarray}

%\newpage
%%%%%%%%%%%%%%%%%%%%%%%%%%%%%%%%%%%%%%%%%%%%%%%%%%%%%%

\end{document}